\documentclass[11pt]{article}

\usepackage[latin1]{inputenc}
\usepackage{eulervm}
\usepackage[top=1.1in, bottom=1.1in, left=1.15in, right=1.15in]{geometry}
\usepackage{setspace}
\usepackage[pdftex]{graphics}
\usepackage{graphicx}
\usepackage{psfrag}
\usepackage{color}
\usepackage{latexsym}
\usepackage{moreverb} 
\usepackage{amsfonts}
\usepackage{amssymb, amsmath}
\usepackage{amsthm}
\usepackage{subcaption}
\usepackage{float}
\usepackage{bm}
\usepackage{har2nat}
\usepackage{enumitem}
\usepackage{tikz}
\usepackage{natbib}
\usepackage[colorlinks=true, pdfstartview=FitV, linkcolor=blue, citecolor=blue, urlcolor=blue]{hyperref}
\DeclareGraphicsExtensions{.pdf,.png,.jpg}
\usepackage[commandnameprefix=always]{changes}

\def\I{\mathcal{I}}
\def\E{\textnormal{E}}

\def\Q{\textnormal{Q}_{\tau }}

\def\R{\mathbb{R}}

\def\X{\mathcal{X}}
\def\Y{\mathcal{Y}}

\def\Z{\mathcal{Z}}

\usepackage[normalem]{ulem}

\title{Implicit quantile preferences of the Fed and the Taylor rule}

\author{Gabriel Montes-Rojas\footnote{CONICET and Instituto Interdisciplinario de Econom\'ia Pol\'itica, Universidad de Buenos Aires, Ciudad Aut\'onoma de Buenos Aires, Argentina. gabriel.montes@economicas.uba.ar}  \and Fernando Toledo\footnote{Universidad Nacional de La Plata, La Plata, Argentina. fernando.toledo@econo.unlp.edu.ar}
\and Nicol\'as Bertholet\footnote{Instituto Interdisciplinario de Econom\'ia Pol\'itica, Universidad de Buenos Aires, Ciudad Aut\'onoma de Buenos Aires, Argentina. nicolasbertholet2008@economicas.uba.ar}
\and Kevin Corfield\footnote{Universidad de Buenos Aires, Ciudad Aut\'onoma de Buenos Aires, Argentina. kevincorfield@economicas.uba.ar}
}\date{October 2025}

\begin{document}

\maketitle

\begin{abstract}
We study optimal monetary policy when a central bank maximizes a quantile utility  objective rather than expected utility. In our framework, the central bank's risk attitude is indexed by the quantile index level, providing a transparent mapping between hawkish/dovish stances and attention to adverse macroeconomic realizations. We formulate the infinite-horizon problem using a Bellman equation with the quantile operator. Implementing an Euler-equation approach, we get Taylor-rule-type reaction functions.  Using an indirect inference approach, we derive a central bank risk aversion implicit quantile index. An empirical implementation for the US is outlined based on reduced-form laws of motion with conditional heteroskedasticity, enabling estimation of the new monetary policy rule and its dependence on the Fed risk attitudes. The results reveal that the Fed has mostly a dovish-type behavior but with some periods of hawkish attitudes.
\vspace{7mm}

\vspace{3mm}

\end{abstract}

\newpage

\section{Introduction}

The debate between hawkish and dovish monetary policy authorities has received increased attention (see  \citeasnoun{tnm17}, \citeasnoun{hkm2025}). For example, the London Investment Service Company In Touch Capital Markets has introduced a relative scale to explain the position of Fed's regarding their relative aversion to inflationary pressures: the lowest bound on the scale indicates very hawkish members and the upper bound specifies very dovish ones. In the empirical literature, CBs are often classified as hawkish or dovish based on projected Taylor rules, particularly the interest rate's long-term response to inflation (see \citeasnoun{castro11}; \citeasnoun{wilson20}; \citeasnoun{malmendier21}; \citeasnoun{gonzalezandtanvir23}).

The responses of monetary authorities to the post-pandemic period and the Ukraine war are paramount to understand coordinated monetary policy actions in terms of interest rate hikes to recent global inflation synchronization (\citeasnoun{haglobalinflation24}). Although this empirical regularity, the intensity of restrictive monetary policy measures differs among countries. 
This stylized fact reminds us how important it is for CBs to find an optimal policy monetary rule that reflects the distinct degree of risk aversion to inflation and economic fluctuations. For example, the Fed's approach was shaped by empirical evidence that gathered in the years before the pandemic, and the results were impacted by a decrease in inflation persistence, a flattening of the Phillips curve's slope, and inaccurate assessments of real-time output or unemployment (\citeasnoun{sargentargentandwilliams25}). 

Taylor rules have become a cornerstone of modern monetary policy science (\citeasnoun{woodford2003interest}, \citeasnoun{jbt93}). Optimal monetary policy involves setting short-term nominal interest rates to stabilize the economy by managing inflation and output gaps, often guided by rules that recommend increasing rates when inflation is high or output exceeds its potential and lowering them otherwise (\citeasnoun{ClaridaGaliGertler99}).

Although Taylor rules provide a simple and robust framework, true optimal policy depends on specific economic models, policy objectives, and whether policymakers prioritize inflation, output, or a combination of both, leading to ongoing debates and refinements of such rules. In that regard, during the last Jackson Hole Symposium (August 2025, 21-23), \citeasnoun{nakamuraribliersteinsson2025} presented a study that remarks the descriptive nature of Taylor rules instead of their prescriptive one. These authors have also observed deviations from the Taylor principle after exploring the recent Fed's behavior and also pointed out the coexistence of early and late policy interest rate hikers.  

In the case of New Keynesian macroeconomic models, the discussion about the theoretical validity of optimal Taylor rules usually includes these and other relevant issues (see \citeasnoun{bh2014}). The main point we would like to state here is the discontent that some contributions express in terms of the theoretical form of optimal Taylor rules (see \citeasnoun{coch2007}, \citeasnoun{bsgu2001}).  

Theoretical contributions have usually approached the different preferences toward the traditional monetary policy trade-off (inflation versus output fluctuations) by deriving an optimally monetary policy Taylor rule minimizing a quadratic loss intertemporal function (\citeasnoun{woodford2003interest}). However, problems arise in applying these functions when the underlying economic model is nonlinear (\citeasnoun{be2023}), shocks are non-normal (\citeasnoun{hmm2024}), or the quadratic assumption is too simplistic (\citeasnoun{ns2002}). Furthermore, although linear-quadratic models offer analytical tractability for deriving optimal rules, they may fail to capture key asymmetric preferences that drive actual monetary policy behavior in the real world (\citeasnoun{svensson2003})\footnote{\citeasnoun{el-shagi25} has recently shown that the Fed prioritizes business cycle stabilization over containing inflation.}. So, traditional methods often look at the average policy reaction function of a CB. 

After the 2008 global financial crisis, many authors argued that CBs should pay more attention to tail risks (especially downside risks) rather than just average outcomes (see \citeasnoun{demirguc2013}). For instance, regardless of the US economy's condition, \citeasnoun{barci25} notes that monetary policy can increase downside risk; nevertheless, this ability is significantly diminished during economic expansions. Such disparity, if not appropriately taken into consideration, could cause monetary authorities to be too cautious when it comes to tightening during booms. 

In the present paper, our theoretical contribution is to introduce quantile utility (QU) preferences into the intertemporal minimization of the loss function of CBs. As far as we know, this contribution is completely new. We add to the literature on quantile preferences applications a new analytical framework to analyze how CBs minimize their intertemporal loss function using dynamic programming through the Bellman equation solution as in \citeasnoun{deCastroGalvao19}. In contrast to the usual framework of minimizing a quadratic intertemporal loss function using expected utility to attain the optimal policy Taylor rule, we incorporate QU preferences to study how CBs respond to undesirable macroeconomic outcomes. It is worth noting that our analytical framework assumes full credibility of the policymaker's announcements (see \citeasnoun{woodford2003interest}). 


Our research adds new arguments to the theoretical discussion about the accurate form of the optimal Taylor rule. We depart from the conventional dynamic optimization problem faced by CBs to propose a new closed analytical form for their reaction function. The main advantage of the new Taylor rule is that CBs are concerned not only with average inflation-output trade-off but also in analyzing scenarios of inflation and output gaps. 

Moreover, by comparing the observed policy actions (i.e. interest rate) with the entire myriad of available actions for all quantiles, we can infer the type of the CB at each point in time. We define this the implicit quantile preference index, which is a dynamic index of the CB risk aversion. In turn, we intepret this index as a dovish/hawkish scale.

We use US quarterly long-run data from 1954-Q4 to 2025-Q2. This framework suggests that the Fed often appear more dovish than simple Taylor rules would suggest and they may be intrinsically less concerned with downside risks (unemployment spikes, financial instability). However, this is not a general description, and there are specific periods where the Fed is characterized with hawkish behavior. Our analysis delivers an index of the Fed's risk aversion attitudes across time that is based on the implicit quantile preference.

This study relates to three branches of the literature. First, we add to the literature on optimal monetary policy the idea that CBs adjust their monetary policy actions to more complex optimal rules than traditional ones. We show how our theoretical framework relates to the New Keynesian macroeconomic models by getting a new Taylor rule that allows different risk aversion attitudes towards inflation and output combinations. Second, we contribute to the quantile preferences literature (see \citeasnoun{deCastroGalvao22}) with an innovative application: the formal derivation of a closed form for a new Taylor rule. We use dynamic programming methods and the Bellman equation to optimize the CB intertemporal loss function and obtain the new optimal reaction function of policymakers (see \citeasnoun{deCastroGalvao19}; \citeasnoun{hills2020}). Third, we deliver an indirect inference approach to estimate the Fed's risk attitude across time, thus contributing to the analysis of parameter and/or model uncertainty (see \citeasnoun{cogley2008}).  

It should be noted that our approach is different from empirical papers that estimate heterogeneous responses in a Taylor rule regression model as in \citeasnoun{Chevapatrakul2009}, \citeasnoun{wolters2012estimating}, \citeasnoun{chevapatrakul2014monetary}, \citeasnoun{chen2017japanese} and \citeasnoun{Christou2018}, among others. In those papers, the key goal is to evaluate heterogeneous responses of the interest rate to inflation and output gap (and others) using a quantile regression framework. In our paper, quantiles relate to a structural preference parameter of the CB and not to the conditional quantiles of the conditional distribution of the interest rate.

The paper proceeds as follows. Section \ref{sec:QU} summarizes the quantile utility framework. Section \ref{sec:taylorrule} applies dynamic programming to the intertemporal QU maximization problem to derive Taylor rules. Section \ref{sec:empiricalmodel} describes the empirical implementation strategy. Section \ref{sec:econ model results} presents the estimation results. Section \ref{sec:conclusions} concludes.

\section{Quantile preferences and risk attitude}\label{sec:QU}

\subsection{Quantile preferences for univariate outcomes}

An expected utility (EU) maximizer with utility function $u:\R \to \R$ prefers lottery $X$ to $Y$  if $\E[u(X)] \geqslant \E[u(Y)]$. This refers to a case when a decision maker (DM) that is faced with uncertain outcomes chooses the action that maximizes the expected average outcome. Quantile utility\footnote{
Quantile preferences were first introduced by \citet{Manski88}. \citeasnoun{Rostek:10} and 
\citeasnoun{Chambers:09} provide axioms for the static case, and \citeasnoun{deCastroGalvao22} formally axiomatize both the static and dynamic quantile preferences. 
\citeasnoun{Giovannetti13} studies a two-period economy for an intertemporal consumption model under quantile utility maximization.
\citeasnoun{deCastroGalvao19}  establish the properties of a 
general dynamically consistent quantile preferences model. We refer to this type of preference modeling as quantile utility (QU hereafter).} (QU) is based on a framework where optimal decisions and allocations correspond to maximizing a specific quantile of the distribution of outcomes or returns.\footnote{
Given a random (univariate) variables, $Y$, let $F(y) = F_{Y} (y) = \Pr \left( Y \leqslant y \right)$ denote the  conditional cumulative distribution function (c.d.f.) of $Y$. 
If the function $y \mapsto F_{Y} (y)$ is strictly increasing and continuous in its support, its inverse is the quantile of $Y$, that is, $\Q[Y] = F_{Y}^{-1} (\tau) $, for $\tau \in (0,1)$.
If $y \mapsto F_{Y} (y)$ is not invertible, 
we can still define the quantile as one of its  generalized inverses. 
Following the standard practice, we define the quantile as the left-continuous version of the generalized inverse:%
\begin{equation*}
    \Q [Y] \equiv \inf \{y \in \R : \Pr[Y  \leqslant y ] \geqslant \tau\}. 
\end{equation*}
 } Quantile preferences are defined by simply substituting the expectation by the quantile operator, that is, 
\begin{align}\label{eq:def_Q_with_U}
X \succeq Y &\iff \Q [u(X)] \geqslant \Q [u(Y)]. 
\end{align}
The intuition is that, in the presence of uncertainty, a QU maximizer makes decisions based on maximizing a given $\tau$ quantile of the distribution of potential outcomes.


For univariate random variables (i.e. monetary outcomes), quantiles enjoy the following property: for any continuous and  increasing function $f: \R \to \R$, $f( \Q[X]) = \Q [ f(X)]$.
If   $u:\R \to \R$ is strictly increasing and continuous, as usual, then we can take its inverse and apply to \eqref{eq:def_Q_with_U}, to obtain: 
\begin{align*}
X \succeq Y &\iff u^{-1} ( \Q [u(X)] )  \geqslant 
u^{-1} (
\Q [u(Y)] )  \iff  \Q [X] \geqslant \Q [Y] .
\end{align*}
In other words, for the univariate case, the utility functional form is not necessary to model DM behavior.

As noted by \citeasnoun{Manski88}, under QU risk attitudes can be indexed by $\tau$ itself.  Intuitively, we can map risk aversion into the $\tau$ scale, such that a $\tau$-DM is more risk averse than $\tau'$-DM if $\tau\leq\tau'$. In sum, a $\tau$-DM evaluates lotteries and actions based on choosing the ones with the highest $\tau$ quantile, and then, the lower $\tau$ is, the more the DM is concerned with low values or losses.

\subsection{Quantile preferences for multivariate outcomes}

For a multivariate random outcome variable, say $m$ dimensional vectors of the form \( Y =(Y_1,Y_2,...,Y_m)\) with domain \( \Y \subseteq \mathbb{R}^m \), \( \inf \{ y \in \mathcal{Y} : \tau \leq F(y) \} \) is (in general) not unique.

\begin{figure}[ht]
    \centering
    \caption{(a) Two points representing quantile $\tau$. $P(Y_1 \leq q_1, Y_2  \leq q_2) =\tau$. (b) Two contour lines for $\tau<\tau'$}
    \label{fig:biv} 
    \centering
    \begin{tabular}{cc}
    (a) & (b) \\
        \includegraphics[scale=0.6]{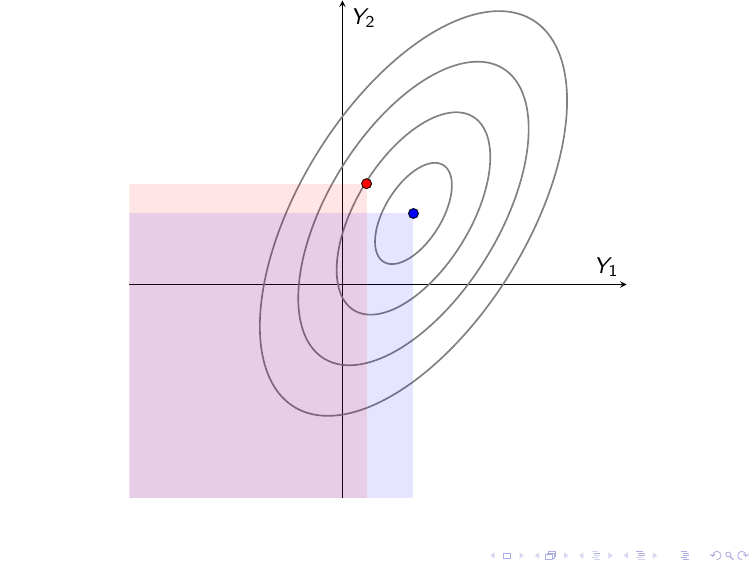}&
        \includegraphics[scale=0.6]{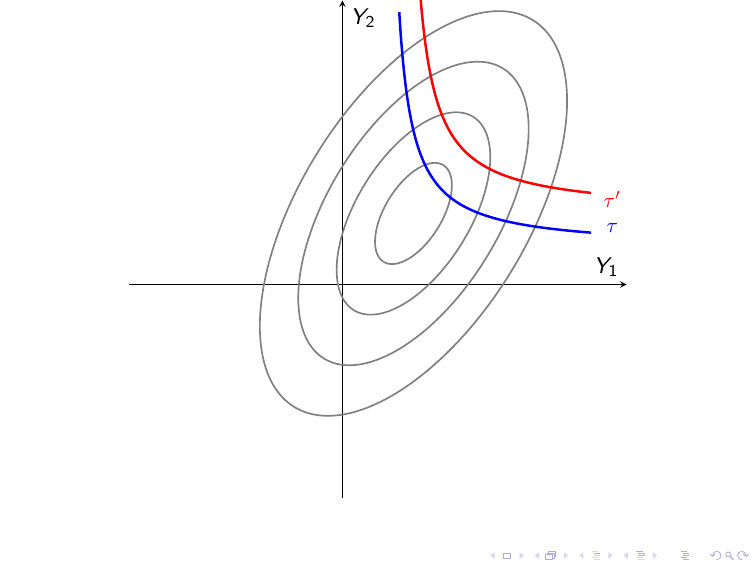}
    \end{tabular}
\end{figure}

Take the bivariate case \( Y =(Y_1,Y_2)\) with domain \( \mathcal{Y} \subseteq \mathbb{R}^2 \). Quantiles are themselves then defined on regions, contours and depths \citep{HallinKonen24}.
A quantile on the bivariate domain is any pair $(q_1,q_2)\in\Y$ such that $P(Y_1 \leq q_1, Y_2 \leq q_2)=\tau$, $\tau\in(0,1)$. 
Figure \ref{fig:biv}a plots the contour plot for the probability density function of a bivariate distribution and adds two points that correspond to the same quantile $\tau$. Figure \ref{fig:biv}b plots the same contour plot but considers the curves corresponding to two different $\tau$'s.

\begin{figure}[ht]
    \centering
    \caption{(a) $Q_\tau(X)>>Q_\tau(Y)$. (b) $Q_\tau(X)><Q_\tau(Y)$}
    \label{fig:Q2} 
    \centering
    \begin{tabular}{cc}
    (a) & (b) \\
        \includegraphics[scale=0.6]{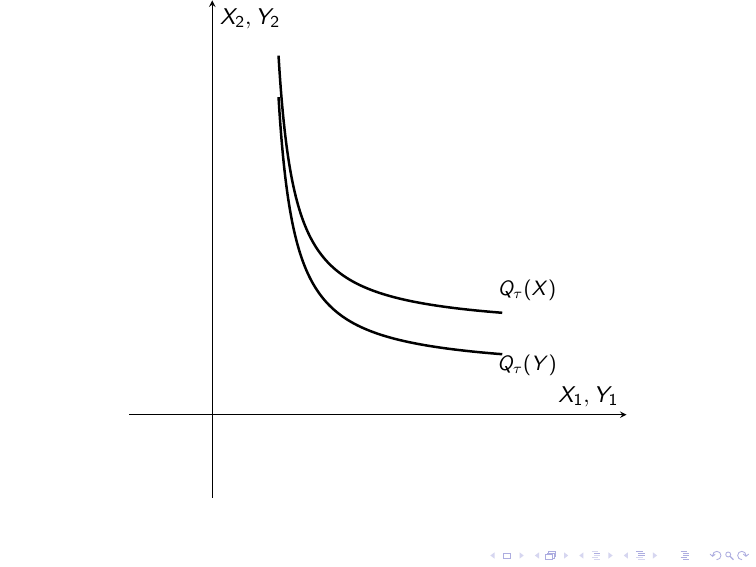}&
        \includegraphics[scale=0.6]{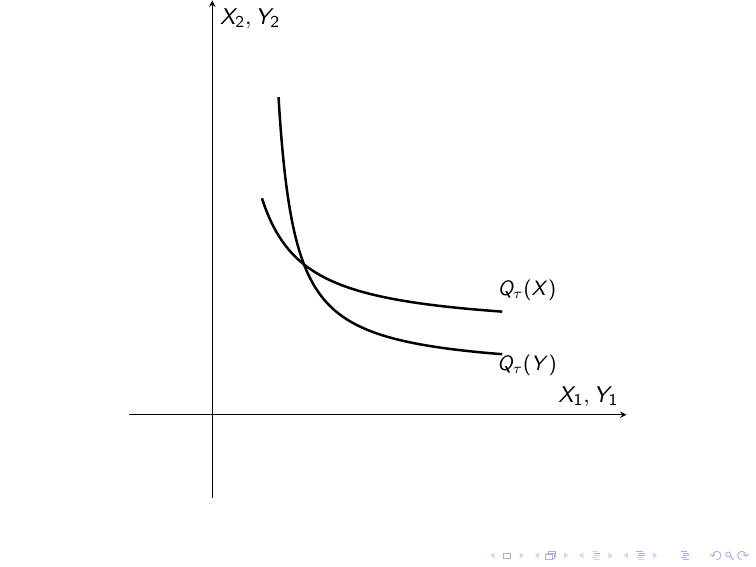}
    \end{tabular}
\end{figure}

Note that the preceding analysis of QU cannot be applied to the multivariate domain unless additional considerations are taken. Consider two random variables $X=(X_1,X_2)$ and $Y=(Y_1,Y_2)$ on the bivariate domain. Figure \ref{fig:Q2} plots two different cases for the same quantile $\tau$ with (a) and without (b) crossing. As such, there is no natural ordering that can be used in terms of the distribution function or its inverse, the quantiles.

A consequence of this is that the QU model cannot evaluate random utility based on the multivariate distribution of the arguments determining the utility. For our purposes, a QU-maximizer CB that has preferences over inflation and output gap cannot resort to the joint distribution of these variables to evaluate policies. On the contrary, it does require the utility function and the relative valuation of each component.

Following \citet{HPS10} multivariate models can be decomposed into a series of univariate models in terms of quantile analysis. Quantiles are analyzed in terms of a \textit{magnitude} and a \textit{direction}. We define 
\(T = (\tau_1, \tau_2, \ldots, \tau_m) \in (0,1)^m\) as a set of quantile indices. 
The vector \(T\) can be factorized as \(T \equiv \tau v\), where 
\(\tau = \|T\| \in (0,1)\) represents the \textit{magnitude}, and 
\(d \in \mathbb{R}^{m-1} \equiv \{d \in \mathbb{R}^m : \|d\| = 1\}\) 
represents the \textit{direction} expressed as a unit vector in the Euclidean framework. 

In this model, \(\tau\) is a scalar quantile index that specifies the position along the distribution, 
while \(\gamma\) is a unit vector that determines the direction in the \(m\)-dimensional space. 
This vector can be interpreted as an \((m - 1)\)-dimensional directional component that captures how quantile changes unfold across variables. This decomposition allows for an intuitive and geometric interpretation of multivariate quantiles in terms of distance and orientation within the variable space. 

\begin{figure}
\centering
    \caption{Vector directional quantile}
    \label{fig:DQ}
    \includegraphics[width=8cm]{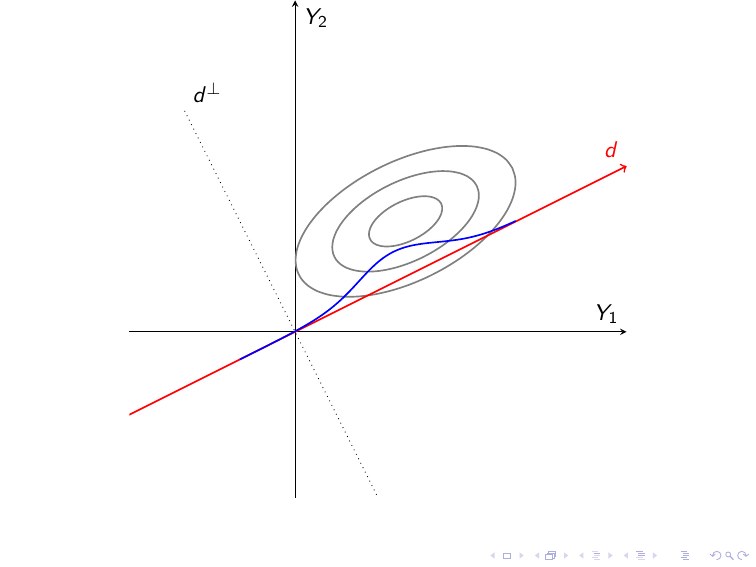}
\end{figure}

Vector directional quantile proposes to study univariate variables of the form $d\cdot y$, where $\cdot$ represent element-by-element vector multiplication. Figure \ref{fig:DQ} plots this idea for the bivariate case (in the figure \(d^\perp\) is an orthonormal basis of the subspace orthogonal to \(d\)). Once a direction is fixed, the problem becomes one of univariate quantiles, and QU analysis can be applied. 

In the QU setup, the direction $d$ can be interpreted as a linearization of utility function over the multivariate domain. The direction reduces the dimensionality of preferences into a linear univariate model. Note, however, that is only valid in a local sense. Non-local comparisons require the use of the utility function to fulfill this role for all cases. 

\subsection{Dynamic models}

Many applications of intertemporal maximization use the standard recursive EU. 
These models have been workhorses in several economic fields.  
EU is simple and amenable to theoretical modeling. 
The assumption of maximization of average utility, the average being a simple measure of centrality, has intuitive appeal as a behavioral postulate. 
Nevertheless, the usual EU framework has been subjected to a number of criticisms, including in its dynamic version. For example, it has been well documented in the literature that it is not possible to separate the intertemporal substitution from the risk attitude parameters when using standard dynamic models based on the EU \citep[see, e.g., ][]{Hall88}. 
The framework proposed by  \citet{KrepsPorteus:78} to study temporal resolution of uncertainty was one of the first efforts to go beyond EU in the dynamic setting. 
An expanding literature considers alternative recursive models.\footnote{
We refer the reader to  \citet{EpsteinZin89,EpsteinZin91}, \citeasnoun{Weil:90}, \citeasnoun{GrantKajiiPolak00}, \citeasnoun{EpsteinSchneider:03a}, \citeasnoun{HansenSargent04}, \citeasnoun{MaccheroniMarinacciRustichini:06},  \citeasnoun{KlibanoffMarinacciMukerji:09}, \citeasnoun{MarinacciMontrucchio10}, 
\citeasnoun{Strzalecki:13}, 
\citeasnoun{BommierKochovLeGrand17}, \citeasnoun{Sarver18}, and \citeasnoun{DeJarnetteDillenbergerGottliebOrtoleva2020} among others.}

\citeasnoun{deCastroGalvao19} developed a new alternative to the EU recursive model based on QU. In their model, the economic agent chooses the  alternative that leads to the the highest $\tau$-quantile of the stream of future utilities for a fixed $\tau \in (0,1)$.
The dynamic quantile preferences for intertemporal decisions are represented by an additively separable quantile model with standard discounting. 
The associated recursive equation is characterized by the sum of the current period utility function and the discounted value of the certainty equivalent, which is obtained from a quantile operator. 
This intertemporal model is tractable  and simple to interpret, since the value function and Euler equation are transparent, and easy to calculate (analytically or numerically). 
This framework allows for the separation of the risk attitude from the intertemporal substitution, which is not possible with EU, while maintaining important features of the standard model, such as dynamic consistency and monotonicity.

\section{Preferences of a CB and Taylor rule for QU maximizer}\label{sec:taylorrule}

\subsection{General set-up}

Consider now a CB that has $\tau$-QU preferences  based on $(\pi_t-\pi^*)$ where $\pi_t$ is inflation and $\pi*$ is the target inflation rate, and output gap $y_t=(\bar y_t-y_t^*)$ where $\bar y_t$ is output and $y_t^*$ is a measure of output long-run trend and potential output. Moreover, we assume that the CB has a preference for smoothing policy variables over time (i.e. the interest rate).

The CB decisions can be represented along a utility function $u(y,\pi,i,z)$ that typically is a trade-off between inflation and output gaps, and it may depend on interest rate and current shocks.
In general, we could assume that the CB values more inflation and output closer to the target values. 

We assume a quadratic utility function of the form
\begin{equation}\label{eq:lossfunction}
    u(\pi_t,y_t,i_t,z_t)=-\frac{(\pi_t-\pi^*)^2}{2}-\frac{\lambda (y_t)^2}{2}-\frac{\delta (i_t-i_{t-1})^2}{2}.
\end{equation}
In this model, the CB has preferences for state variables close to the target values and for avoiding fluctuations in the interest rate.
This utility function is in fact a loss function multiplied by $-1$, and it can be derived from micro-foundations as in \citeasnoun{woodford2003interest}. $\lambda$ and $\delta$ are structural parameters that correspond to the degree of substitution of the inflation gap, the output gap and the variations in interest rate along indiference curves.

Let $x\in\mathcal{X}$ denote the particular state and the state space, $i\in\mathcal{I}$ be the action and the set of possible actions the CB may take, and $z\in\mathcal{Z}$, the range of the shocks.  For our purposes, $x_t=(y_t,\pi_t)$, $i_t$ is the interest rate and $z_t$ represents random components that affect $x_t$. Moreover, we consider that $z_t=(z_{\pi,t},z_{y,t})$ a bivariate random vector. Although we do not explicitly consider it to reduce notation, the state variables may include lags of the variables. For our particular case, we use the lag of $i$ inside the utility function.

The next period state, $x_{t+1}$, is defined by a law of motion function $\phi : \mathcal{X} \times \mathcal{I} \times \mathcal{Z} \rightarrow \mathcal{X}$ that satisfies $x_{t+1}=\phi(x_t,i_t,z_{t+1})$.
Given the current state  $x_t$ and current shock $z_t$,  $\Gamma (x_t,z_t)$ denotes the  set of possible choices $i_{t}$, that is, the feasibility constraint set.  

\subsection{Infinite horizon and recursive maximization problem}

In the proposed CB model, the uncertainty with respect to the future realizations of $z_{t}$ is given by a quantile applied to potential values of the utility function. In line with QU theory, the quantile index $\tau$ represents CB attitudes towards risk. We refer to a $\tau'$-CB to be more risk averse than a $\tau$-CB one if $\tau'<\tau$. That is, the $\tau'$-CB is more concerned with worse outcomes scenarios (i.e. high inflation, low output) than a $\tau$-CB. 

In the QU framework, optimal decisions are taken to maximize the $\tau$ quantile of intertemporal utility in an infinite horizon problem. 
This framework does not allow for the same solution strategies as in the EU case, because we cannot apply the law of iterated expectations. However, under certain conditions described in \citeasnoun{deCastroGalvao19}, these dynamic intertemporal choices can be represented by the maximization of a value function $v : \X \times \Z \to \R$ that satisfies the recursive Bellman equation: 

\begin{equation}\label{eq:FE}
v(x,z)= 
\sup_{
i \in \Gamma (x,z) }
\; \; 
\Bigl\{  u\left(x, i , z \right) \; + \;  \beta\Q  [  \; v \left( \phi(x,i,z') , \, z' \right) \; | \; z] \Bigr\},
\end{equation}
where $z'$ indicates the next period shock. 

Note that this is similar to the usual dynamic programming problem, in which the expectation operator $\E[\cdot]$  is in place of $\Q [\cdot]$. 
\citeasnoun{deCastroGalvao19} and \citeasnoun{deCastroGalvaoNunes25} endorse the construction of this type of recursive models from dated preferences. Those authors prove uniqueness of the solution to problem \eqref{eq:FE}, under a set of regularity conditions similar to those in dynamic programming set-up and some specific restrictions for the use of quantiles. The solution is a policy function $i_\tau^{\ast}: \X \times \Z \to \I$, that associates to each  $(x_t, z_t)$ the optimal solution  $i_\tau^\ast = i^{\ast}(x_t,z_t)$. 

\subsection{Law of motion}

Now consider a location-scale law of motion $\phi(.)$ for inflation and output gap, using autoregressive processes of order 1. 

\[
\pi_{t+1} = \phi_\pi(x_t,i_t,z_{\pi,t+1}) = \alpha_{\pi0}+\alpha_{\pi\pi} \pi_t + \alpha_{\pi y} y_t + \alpha_{\pi i} i_t + h_\pi(\pi_t,y_t,i_t)z_{\pi,t+1},
\]
\[
y_{t+1} = \phi_y(x_t,i_t,z_{y,t+1}) = \alpha_{y0} + \alpha_{y\pi} \pi_t + \alpha_{yy} y_t + \alpha_{y i} i_t + h_y(\pi_t,y_t,i_t)z_{y,t+1},
\]
where the $\alpha$ coefficients capture location mean effects, i.e. the persistence of inflation and output gap and how sensitive inflation and output gap are to changes in the interest rate \( i_t \), and \( z_{\pi,t+1} \) and \( z_{y,t+1} \) are random shocks, possibly correlated to each other but assumed to be independent of the state and interest rate variable, i.e. $(z_{\pi,t+1},z_{y,t+1})\perp  (\pi_t,y_t,i_t)\mid z_t$. They
have zero conditional mean $E[z_{a,t+1}|z_{\pi,t},z_{y,t}]=0,\ a=\pi,y$ and unit variance $E[z_{a,t+1}^2|z_{\pi,t},z_{y,t}]=1$ (which anyway cannot be identified separately from $h$). This is a reduced form that may be the result of intertemporal IS curve and a New Keynesian Phillips curve as in \citeasnoun{woodford2003interest} and \citeasnoun{gali2015monetary}. 



Functions $h_\pi(.)$ and $h_y(.)$ are skedastic strictly positive functions that control the conditional heteroskedasticity of the state variables, affecting the scale. In turn, they determine the structure of heterogeneity in the law of motion and whether the quantiles  are not parallel to each other. We can refer to location shift only models to those where the $h$ functions are constant, and to location-scale shift models where the $h$ functions depend on $(\pi,y,i)$.

Several parameterizations can be applied, see for instance \citeasnoun{RomanoWolf17}. Different specifications used in the heteroskedasticity literature to model location-scale shift effects are

\[
h_a(\pi_t,y_t,i_t)=(\gamma_{a0}+\gamma_{a\pi}\pi_t+\gamma_{a y}y_t+\gamma_{a i}i_t)^{1/2},\ a=\pi,y.
\]
or
\[
h_a(\pi_t,y_t,i_t)=exp\left[(\gamma_{a0}+\gamma_{a\pi}\pi_t+\gamma_{a y}y_t+\gamma_{a i}i_t)^{1/2}\right],\ a=\pi,y.
\]
The $\gamma$ coefficients control whether the random shocks affect the scale impact of these shocks on inflation and output gap. Thus models with $\gamma=0$ have only location shifts, while $\gamma\neq0$ characterize location-scale shift ones. 

In the context of quantile regression specifications, this representation allows for a random-coefficient model indexed by a quantile index, i.e.,

\[
Q_{\tau_\pi}(\pi_{t+1}|x_{t},i_t) = \alpha_{\pi0}(\tau_\pi)+\alpha_{\pi\pi}(\tau_\pi) \pi_t + \alpha_{\pi y}(\tau_\pi)y_t + \alpha_{\pi i}(\tau_\pi)i_t, \ \tau_\pi\in(0,1),
\]
\[
Q_{\tau_y}(y_{t+1}|x_{t},i_t) = \alpha_{y0}(\tau_y)+\alpha_{y\pi}(\tau_y) \pi_t + \alpha_{yy}(\tau_y) y_t + \alpha_{y i}(\tau_y) i_t, \ \tau_y\in(0,1),
\]
where $\alpha_{ab}(\tau_a)=\alpha_{ab}+\frac{\partial h_a(x_{t},i_t)}{\partial b}Q_{\tau_a}(z_{a}|x_{t},i_t)$, $a=\pi,y$ and $b=0,\pi,y,i$. Here $\tau_\pi$ and $\tau_y$ reflects different conditional responses of inflation and output gap to current state variables and policy choices. Both indexes are not necessarily independent and they need to be considered in a multivariate quantile model as in \citeasnoun{Montes17,Montes19,Montes22}.

 \subsection{Euler equations}

Under certain conditions, the Taylor rule can be derived analytically from this function by implementing the Euler equation as in \citeasnoun{deCastroGalvaoNunes25} Theorem 3.18. Consider an application of the theorem to get the Euler equation as:

\begin{eqnarray}\label{eq:Euler}
\frac{\partial u(x_t,i_t,z_t)}{\partial i} + &\nonumber\\
\;  \beta\Q  \left[  \frac{u(x_{t+1},i_{t+1},z_{t+1})}{\partial \pi} \frac{\partial \phi_\pi(x_t,i_t,z_{t+1})}{\partial i}
+\right. &\nonumber\\
\left.\left.\frac{u(x_{t+1},i_{t+1},z_{t+1})}{\partial y} \frac{\partial \phi_y(x_t,i_t,z_{t+1})}{\partial i}
\right| \; z \right]&=0.
\end{eqnarray}
For this derivation to be applied, it requires that differentiability and the quantile operator can be interchanged, and that the shocks have an increasing monotonic effect. 
In particular, we need the following univariate component 

$$\frac{u(x_{t+1},i_{t+1},z_{t+1})}{\partial \pi} \frac{\partial \phi_\pi(x_t,i_t,z_{t+1})}{\partial i}+\frac{u(x_{t+1},i_{t+1},z_{t+1})}{\partial y} \frac{\partial \phi_y(x_t,i_t,z_{t+1})}{\partial i}$$
to be monotonically increasing on $z$. Note, however, that $z$ is bivariate, and therefore the monotonicity requirement has to be evaluated at particular vector directions. For the case in eq. \eqref{eq:lossfunction},
\[\frac{u(x_{t+1},i_{t+1},z_{t+1})}{\partial \pi}=-(\pi_{t+1}-\pi^*),\]
\[\frac{u(x_{t+1},i_{t+1},z_{t+1})}{\partial y}=-\lambda y_{t+1},\]
\[\frac{u(x_{t},i_{t},z_{t})}{\partial i}=-\delta (i_t-i_{t-1}).\]
Then the monotonicity requirement is that the random variable $q(x,i):=-\phi'_{\pi,i}(x,i)z_\pi-\lambda\phi'_{\pi,i}(x,i)z_y$ with 
$\phi'_{\pi,i}(x,i)=\frac{\partial \phi_\pi(x,i)}{\partial i}$ and $\phi'_{y,i}=\frac{\partial \phi_y(x,i)}{\partial i}$, has a well defined quantile function.

We derive here Euler equations solutions. For simplicity we assume that $\alpha_{\pi0}=\alpha_{y0}=0$ (but this is not assumed in the empirical application).
 
\subsubsection*{Location shift only}

Suppose first that $\gamma_{ab}=0\ a,b=\pi,y,i$, that is, the quantiles of the random shocks are not affected by the state variables nor by the control variable.

Then

\begin{eqnarray}
-\delta (i_t-i_{t-1})+
\beta \Q \left[-(\alpha_{\pi\pi} \pi_t + \alpha_{\pi y} y_t + \alpha_{\pi i} i_t + z_{\pi,t+1}-\pi^*)\alpha_{\pi i}\right. & \nonumber \\
\left.-\lambda(\alpha_{y\pi}\pi_t+\alpha_{yy} y_t + \alpha_{yi} i_t + z_{y,t+1}) \alpha_{yi}\mid z_t\right]&=0.
\end{eqnarray}
Note that by the requirements on the validity of the Euler implementation for QU, $\Q(-z_{\pi,t+1}\alpha_{\pi i}-\lambda z_{y,t+1}\alpha_{yi}\mid z_t)$ needs to be monotonically increasing in both components $(z_{\pi,t+1},z_{y,t+1})$.

Thus we obtain the following Taylor rule for the $\tau$-QU problem

\[
i^*_\tau(\pi_t,y_t,i_{t-1})=\]
\[
(\delta+\beta(\alpha_{\pi i}^2+\lambda\alpha_{y i}^2))^{-1}\times \left\{\delta i_{t-1}\right.\]
\[-\beta\left[(\alpha_{\pi\pi}\alpha_{\pi i}+\alpha_{y\pi}\alpha_{y i}) \pi_t + \lambda(\alpha_{\pi y}\alpha_{\pi i}+\alpha_{y y}\alpha_{y i}) y_t-\alpha_{\pi i}\pi^*\right]\]
\[\left.+\beta\Q(-z_{\pi,t+1}\alpha_{\pi i}-\lambda z_{y,t+1}\alpha_{yi}\mid z_t)\right\}.\]

This is similar to the typical Taylor rule derivation as in \citeasnoun{gw2003}. In the standard model, since the expectation of the random shocks is zero, the second term becomes zero. 

For QU, however, the quantiles need to be computed on a case-by-case basis. For the location shift case, the quantile index $\tau$ determines the quantiles of the two shocks in a particular direction given by $d_\pi z_{\pi,t+1}+d_yz_{\pi,t+1}$ with $d_\pi=\frac{-\alpha_{\pi i}}{\sqrt{\alpha_\pi^2+\lambda^2\alpha_y^2}}$ and $d_y=\frac{-\lambda\alpha_{y i}}{\sqrt{\alpha_{\pi i}^2+\lambda^2\alpha_{y i}^2}}$.

\subsubsection*{Location-scale on state variables only}

Suppose now that $\gamma_{ai}=0\ a=\pi,y$, that is, the interest rate exerts no scale effect on the random shocks, which may be affected by the state variables.

Then

\begin{eqnarray}
-\delta (i_t-i_{t-1})+\beta \Q \left[-(\alpha_{\pi\pi} \pi_t + \alpha_{\pi y} y_t + \alpha_{\pi i} i_t + h_\pi(\pi_t,y_t)z_{\pi,t+1}-\pi^*)\alpha_{\pi i}\right. & \nonumber \\
\left.-\lambda(\alpha_{y\pi}\pi_t+\alpha_{yy} y_t + \alpha_{yi} i_t + h_y(\pi_t,y_t)z_{y,t+1}) \alpha_{yi}\mid z_t \right]&=0,
\end{eqnarray}

\[
i^*_\tau(\pi_t,y_t,i_{t-1})=\]
\[
(\delta+\beta(\alpha_{\pi i}^2+\lambda\alpha_{y i}^2))^{-1}\times \left\{\delta i_{t-1}\right.\]
\[-\beta\left[(\alpha_{\pi\pi}\alpha_{\pi i}+\alpha_{y\pi}\alpha_{y i}) \pi_t + \lambda(\alpha_{\pi y}\alpha_{\pi i}+\alpha_{y y}\alpha_{y i}) y_t-\alpha_{\pi i}\pi^*\right]\]
\[\left.+\beta\Q(-h_\pi(\pi_t,y_t)z_{\pi,t+1}\alpha_{\pi i}-\lambda h_y(\pi_t,y_t)z_{y,t+1}\alpha_{yi}\mid z_t)\right\}.\]

For the location-scale shift case, the quantile index $\tau$ determines the quantiles of the two shocks in a particular direction given by $d_\pi z_{\pi,t+1}+d_yz_{\pi,t+1}$ with $$d_\pi=-h_\pi(\pi_t,y_t)\alpha_{\pi i}\left(\alpha_{\pi i}^2h_\pi(\pi_t,y_t)^2+\lambda^2h_y(\pi_t,y_t)^2\alpha_{y i}^2\right)^{-1/2}$$ and $$d_y=-\lambda h_y(\pi_t,y_t)\alpha_{y i}\left(\alpha_{\pi i}^2h_\pi(\pi_t,y_t)^2+\lambda^2h_y(\pi_t,y_t)^2\alpha_{y i}^2\right)^{-1/2}.$$ For this case, the direction is state dependent, that is, the QU analysis is $(\pi_t,y_t)$-specific.

\subsubsection*{Location-scale on state and control variables}

Finally, for the general case when there are no restrictions on  $\gamma_{ab},\ a,b=\pi,y,i$ we have

\begin{eqnarray*}
-\delta (i_t-i_{t-1})+ & \nonumber \\
\beta \Q \left[-(\alpha_{\pi\pi} \pi_t + \alpha_{\pi y} y_t + \alpha_{\pi i} i_t + h_\pi(\pi_t,y_t,i_t)z_{\pi,t+1}-\pi^*)(\alpha_{\pi i}+\frac{\partial h_\pi(\pi_t,y_t,i_t)}{\partial i}z_{\pi,t+1})\right. & \nonumber \\
\left.-\lambda(\alpha_{y\pi}\pi_t+\alpha_{yy} y_t + \alpha_{yi} i_t + h_y(\pi_t,y_t,i_t)z_{y,t+1}) (\alpha_{y i}+\frac{\partial h_y(\pi_t,y_t,i_t)}{\partial i}z_{y,t+1})\mid z_t \right]&\nonumber \\
\end{eqnarray*}
\begin{equation}
    =0.
\end{equation}

Here, there is no analytical solution because the quantiles will depend on both $z_a$ and $z_a^2$, $a,b=\pi,y$. (Note that for the expected utility case, the expectation of the square is just replaced by its variance, and thus we could still derive a Taylor rule type model).

\section{Empirical implementation}\label{sec:empiricalmodel}

\subsection{Algorithm for empirical implementation}
Consider time-series data $\{\pi_t,y_t,i_t\}_{t=0}^T$ and set a target value $\pi^*$ and parameters $(\beta,\lambda,\delta)$. Note that we are implicitly defining that the target value for output gap is 0. Define $\mathcal{T}$ as a discrete grid on the interval $(0,1)$.

\begin{enumerate}
    \item Estimate law of motion reduced form VAR(1) models for $(\pi_t,y_t)$ using $i_t$ as an exogenous variable to get the $\alpha$ coefficients.
    \item Estimate the skedastic functions $h_\pi$ and $h_y$ by running reduced form VAR(1) models of squared OLS residuals $\{\hat u^2_{\pi,t+1},\hat u^2_{y,t+1}\}$ on $\{\pi_t,y_t,i_t\}$ to get $\gamma$ coefficients.
    \item Compute $\{\hat z_{\pi,t},\hat z_{y,t}\}_{t=1}^T$ stochastic shocks estimates, i.e. $\hat z_{a,t}=\hat u_{a,t}/\hat h_{a,t}$, $a=\pi,y$. Then compute the empirical quantiles, $\widehat\Q$.
    \item Solve for $i_\tau^*(\pi_t,y_t,i_{t-1})$ for  $\tau\in\mathcal{T}$.
\end{enumerate}

Consider now the evaluation of the underlying preferences of the CB. Here we follow an indirect inference procedure. For each time period $t$, we can evaluate the optimal response for all quantile indexes and then infer the $\tau$ that produces the closest value of the observed policy variable. In other words

\[\hat\tau_t=\arg \min_{\tau\in\mathcal{T}}|i_t-i_\tau^*(\pi_t,y_t,i_{t-1})|.\]

This procedure delivers an index of the implicit risk aversion quantile preferences of the CB.

\subsection{Data sources and model calibration}

\subsubsection{Data}

The empirical estimation involved in step 1 of algorithm is based on three macroeconomic variables constructed from raw data obtained from the Federal Reserve Economic Data of St. Louis (FRED St. Louis) database. The original dataset comprised Real GDP (GDPC1), Potential GDP (GDPPOT), the Effective Federal Funds Rate (FEDFUNDS), and the Personal Consumption Expenditures Chain-type Price Index (PCECTPI). For consistency across series, monthly observations (FEDFUNDS) were converted to quarterly frequency using arithmetic averages. From these sources, we derived the following variables

\begin{enumerate}
    \item Output gap ($y_t$)
        \begin{equation}
            y_t= \bigg(\frac{GDPC1_t}{GDPPOT_t}-1\bigg)*100
        \end{equation}
    \item Inflation ($\pi_t$)
        \begin{equation}
            \pi_t =  100 \times \Delta \ln(PCECTPI_t)
        \end{equation}
    \item Interest rate $(i_t)$, proxied by the Effective Federal Funds rate. 
\end{enumerate}
Table \ref{tab:stats} reports summary statistics of the variables used in this paper.

\begin{table}[ht]
\centering
\caption{Descriptive Statistics} 
\label{tab:stats}
\begin{tabular}{lrrr}
  \hline
Statistic & $i$ & $y$ & $\pi$ \\ 
  \hline
  Mean & 4.62 & -0.27 & 0.78 \\ 
  Minimum & 0.06 & -9.02 & -1.61 \\ 
  1st Quartile & 1.94 & -1.58 & 0.41 \\ 
  Median & 4.33 & -0.20 & 0.66 \\ 
  3rd Quartile & 6.24 & 1.25 & 1.01 \\ 
  Maximum & 17.78 & 5.68 & 2.96 \\ 
   \hline
\end{tabular}
\end{table}

The three constructed series -- the output gap ($y_t$), inflation ($\pi_t$), and the nominal interest rate ($i_t$) -- are employed in step 1 of the empirical algorithm. Specifically, they serve as the input variables for estimating the reduced-form law of motion VAR(1) models, from which the coefficients $\alpha$ are obtained. They are then used in step 2 to compute the skedastic functions and the $z$ shocks components in step 3. 

These estimates provide the foundation for the subsequent steps of the empirical implementation. The final dataset used for estimation spans from the fourth quarter of 1954 to the second quarter of 2025, yielding a total of 283 quarterly observations. 

The second VAR(1) model estimation incorporates two dummy variables to control for the extraordinary shocks associated with major global crises. The first dummy captures the exceptional effects of the global financial crisis, covering the period from the fourth quarter of 2007 to the fourth quarter of 2009. The second dummy accounts for the economic disruptions linked to the COVID-19 pandemic, spanning from the first quarter of 2020 to the first quarter of 2021. Including these variables ensures that the estimated relationships among the models endogenous variables are not biased by these exceptional and exogenous events.

\subsubsection{Calibration}

We calibrate the remaining parameters of the model in step 4 of the algorithm, following the literature on the Taylor rule for the US.
The calibration strategy consists of setting some non-target structural parameters based on empirical evidence.

In particular, following \citeasnoun{dennis2004policy} the policy discount factor is set to $\beta = 0.99$\footnote{This represents the CB's time preference, indicating how much it values future welfare compared to current welfare. A higher discount factor 0.99 means the CB is more patient and cares more about long-term stability.}, the relative weight on output gap stabilization is set to $\lambda = 1$\footnote{The quarterly calibration of lambda for a CB loss function with output equal to 1 involves using the loss function's sensitivity to output deviations to determine the weight (lambda) on output in the loss function, relative to inflation. The value of lambda is adjusted to prioritize output stability and it is a matter of a CB deciding how to weigh output versus inflation.}, and the interest rate smoothing is set to $\delta = 0.1$\footnote{Interest-rate smoothing is the tendency for CBs, including the Fed, to adjust interest rates in small steps over time. A value of delta equal to $\delta = 0.1$ in a policy rule represents the weight on a smoothing term in a simplified model, suggesting that about 10 percent of the adjustment in the desired interest rate is reflected in the policy rate within a quarter, indicating a very gradual policy response. The value of 0.1 is a hypothetical calibration that would imply a very fast adjustment compared to the historical norm (historically estimated to be closer to 0.8 in the US), though still gradual. This behavior can stem from optimal policy choices to reduce volatility and manage expectations, or from practical considerations like market reaction and uncertainty (see \citeasnoun{sackandweiland00}). For our purposes, it helps in evaluating heterogeneity across quantiles in a relatively short period of time.}. Finally, we assume that estimates that the average implied inflation target of the Fed is around 2 percent of annual inflation (0.496 percent in quarterly log differences). Among others, \citeasnoun{Andrade2019} and \citeasnoun{Bianchi2019} calibrate their models using an inflation target consistent with the Fed's 2 percent objective.\footnote{Under former Chair Ben Bernanke, the Fed officially adopted a 2 percent inflation target in January 2012. This move brought the Fed in line with many other central banks and was based on a strategy of price stability, aiming for 2 percent inflation as the longer-run goal for achieving both maximum employment and price stability. Although the 2 percent target was made public and official in 2012, the Fed had been operating with a similar goal behind the scenes since 1996.}


The baseline calibration is summarized in Table~\ref{tab:calibration}.  
    
\begin{table}[H]
    \centering
    \caption{Quarterly calibrated parameters}
    \label{tab:calibration}
    \begin{tabular}{lccc}
        \hline
        \textbf{Parameter} & \textbf{Value} & \textbf{Description} & \textbf{Source} \\
        \hline
        $\beta$     & 0.99   & Policy discount factor & Dennis (2004) \\
        $\lambda$   & 1      & Relative weight on output gap & Dennis (2004) \\
        & & stabilization & \\
        $\delta$    & 0.1   & Interest rate smoothing & Sack and Weiland (2000) \\
        & & parameter & \\
        $\pi^{\ast}$& 0.496   & Quarterly inflation target & Andrade et al. (2019); Bianchi (2019)  \\
        \hline
    \end{tabular}
\end{table}

\section{Results}\label{sec:econ model results}

\subsection{Baseline model}
Tables \ref{tab:var_combined} and   \ref{tab:heteroskedasticity_combined} 
 report the reduced form VAR(1) models and the skedastic functions, respectively. In both cases we use two different models. A baseline model without COVID and Global Financial Crisis (GFC) dummies and another with those dummies included. 

The VAR model reveals that lagged $i$ has a positive effect on inflation but a negative effect on output gap. Note that these estimated parameters do not imply a structural relationship among the variables, but they are only a reduced-form result. 

In a reduced-form VAR model, the sign of the output gap's coefficient in the inflation equation is typically positive, consistent with the New Keynesian Phillips curve. However, this relationship can be obscured, unstable, or even appear with the wrong sign (i.e. the flat Phillips curve puzzle) due to the nature of reduced-form estimation and the influence of other shocks. While the underlying structural relationship is positive, the sign we get from a simple reduced-form VAR estimation is not a reliable estimate of the Phillips curve slope (see \citeasnoun{mavroeidis14}). 

A similar argument applies for understanding the sing of the interest rate on inflation. In that regard, the price puzzle is the empirical finding that interest rate hikes can be followed by rising inflation (see \citeasnoun{sims92}). It is primarily a statistical illusion caused by the econometrician's model failing to account for the fact that the CB is raising rates in anticipation of future inflation. When models are properly specified to include the CB's information, the puzzle usually vanishes, and the standard theoretical relationship holds (see \citeasnoun{stockandwatson01}; \citeasnoun{bernankeandmihov98}; and \citeasnoun{leeperandsimsandzha96}).

In addition, our empirical findings indicate that inflation and output gap have high autoregressive coefficients (close to 0.7 for inflation, 0.9 for output gap). The inclusion of the COVID and the GFC dummies do not change the sign of the estimated coefficients.
 
For the skedastic function, a preliminary analysis (not reported; available upon request) reveals that lagged $i$ is not statistically significant in the skedastic functions, and therefore we impose that $\gamma_{ai}=0\ a=\pi,y$ for the computation of the optimal Taylor rule. In turn, this determines that we follow the location-shift model with control variables only in the skedastic function, and that analytical derivations can be used.

\clearpage
\newpage
\begin{table}[!htbp] \centering 
  \caption{VAR(1) Results Full Sample} 
  \label{tab:var_combined}
\small
\begin{tabular}{@{\extracolsep{5pt}}lcccc} 
\\[-1.8ex]\hline 
\hline \\[-1.8ex] 
 & \multicolumn{2}{c}{Baseline} & \multicolumn{2}{c}{Baseline + dummies} \\ 
\cline{2-3} \cline{4-5}
 & Inflation ($\pi$) & Output gap ($y$) & Inflation ($\pi$) & Output gap ($y$) \\ 
\\[-1.8ex]\hline \\[-1.8ex] 
$i_{t-1}$ & 0.024$^{***}$ & $-$0.029 & 0.024$^{***}$ & $-$0.037$^{*}$ \\ 
 & (0.008) & (0.022) & (0.008) & (0.022) \\ 
$\pi_{t-1}$ & 0.719$^{***}$ & $-$0.130 & 0.718$^{***}$ & $-$0.149 \\ 
 & (0.045) & (0.125) & (0.045) & (0.123) \\ 
$y_{t-1}$ & 0.006 & 0.904$^{***}$ & 0.007 & 0.891$^{***}$ \\ 
 & (0.010) & (0.027) & (0.010) & (0.027) \\ 
Constant & 0.114$^{***}$ & 0.215$^{**}$ & 0.110$^{***}$ & 0.314$^{***}$ \\ 
 & (0.037) & (0.105) & (0.039) & (0.107) \\ 
COVID &  &  & 0.131 & $-$0.922$^{**}$ \\ 
 &  &  & (0.164) & (0.451) \\ 
GFC &  &  & $-$0.025 & $-$0.979$^{***}$ \\ 
 &  &  & (0.123) & (0.337) \\ 
\hline \\[-1.8ex] 
Observations & 282 & 282 & 282 & 282 \\ 
R$^{2}$ & 0.668 & 0.804 & 0.669 & 0.812 \\ 
Adjusted R$^{2}$ & 0.664 & 0.802 & 0.663 & 0.808 \\ 
\hline 
\hline \\[-1.8ex] 
\multicolumn{5}{r}{Note: $^{*}$p$<$0.1; $^{**}$p$<$0.05; $^{***}$p$<$0.01} \\ 
\\ 
\end{tabular} 
\end{table}

\begin{table}[!htbp] \centering 
  \caption{Skedastic Models - Full Sample} 
  \label{tab:heteroskedasticity_combined} 
\small 
\begin{tabular}{@{\extracolsep{5pt}}lcccc} 
\\[-1.8ex]\hline 
\hline \\[-1.8ex] 
 & \multicolumn{2}{c}{Baseline} & \multicolumn{2}{c}{Baseline + dummies} \\ 
\cline{2-3} \cline{4-5}
 & $\hat{u}_\pi^2$ & $\hat{u}_y^2$ & $\hat{u}_\pi^2$ & $\hat{u}_y^2$ \\ 
\\[-1.8ex]\hline \\[-1.8ex] 
$\pi_{t-1}$ & 0.045 & $-$0.311 & 0.078$^{**}$ & 0.101 \\ 
 & (0.040) & (0.463) & (0.038) & (0.337) \\ 
$y_{t-1}$ & $-$0.010 & $-$0.217$^{*}$ & $-$0.001 & $-$0.111 \\ 
 & (0.011) & (0.127) & (0.010) & (0.093) \\ 
COVID &  &  & 0.355$^{**}$ & 19.274$^{***}$ \\ 
 &  &  & (0.176) & (1.568) \\ 
GFC &  &  & 0.775$^{***}$ & 0.305 \\ 
 &  &  & (0.132) & (1.175) \\ 
Constant & 0.087$^{**}$ & 1.153$^{**}$ & 0.033 & 0.467 \\ 
 & (0.041) & (0.464) & (0.039) & (0.345) \\ 
\hline \\[-1.8ex] 
Observations & 281 & 281 & 281 & 281 \\ 
R$^{2}$ & 0.007 & 0.013 & 0.126 & 0.368 \\ 
Adjusted R$^{2}$ & $-$0.0004 & 0.006 & 0.113 & 0.359 \\ 
\hline 
\hline \\[-1.8ex] 
\multicolumn{5}{r}{Note: $^{*}$p$<$0.1; $^{**}$p$<$0.05; $^{***}$p$<$0.01} \\ 
\end{tabular} 
\end{table}

\begin{figure}[ht]
    \centering
    \caption{Baseline results: Taylor interest-rate rule and Fed risk aversion.}
    \label{fig:baseline}
    \begin{subfigure}{0.49\textwidth}
        \centering
        \caption{Observed rate and Taylor rule}
        \label{fig:baseline:tr}        \includegraphics[width=\linewidth]{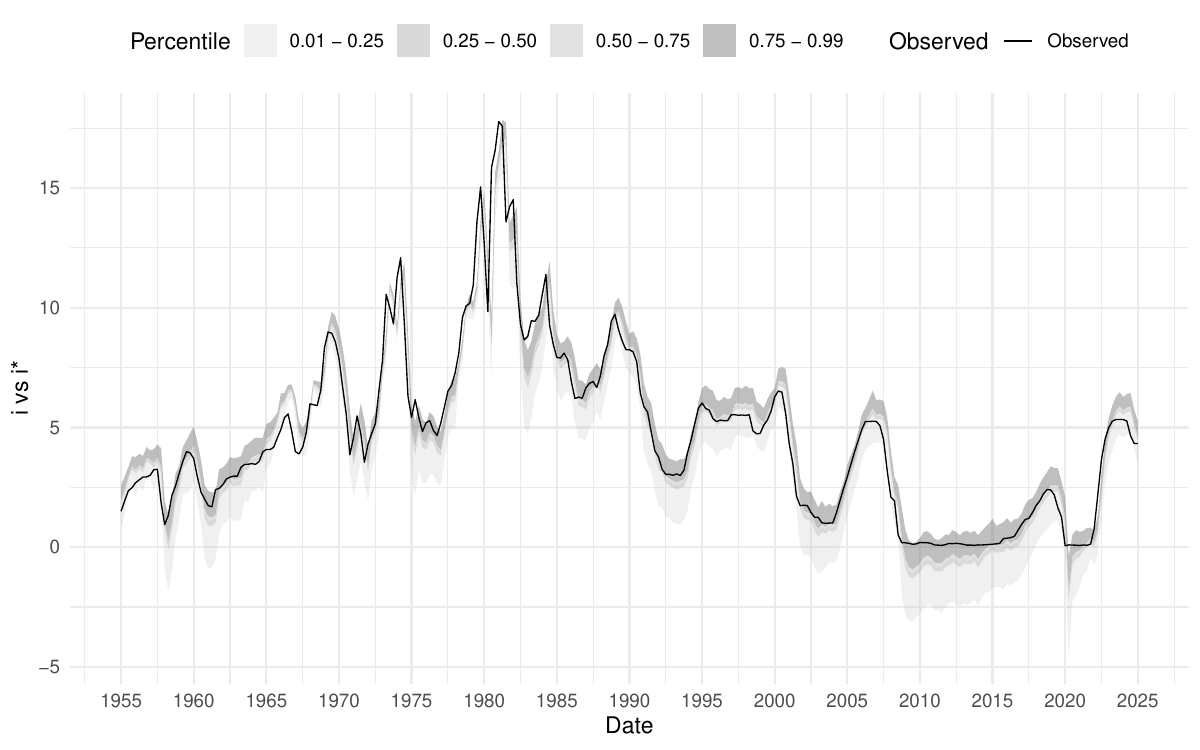}
    \end{subfigure}\hfill
    \begin{subfigure}{0.49\textwidth}
        \centering
        \caption{Implied $\tau$}
        \label{fig:baseline:tau}        \includegraphics[width=\linewidth]{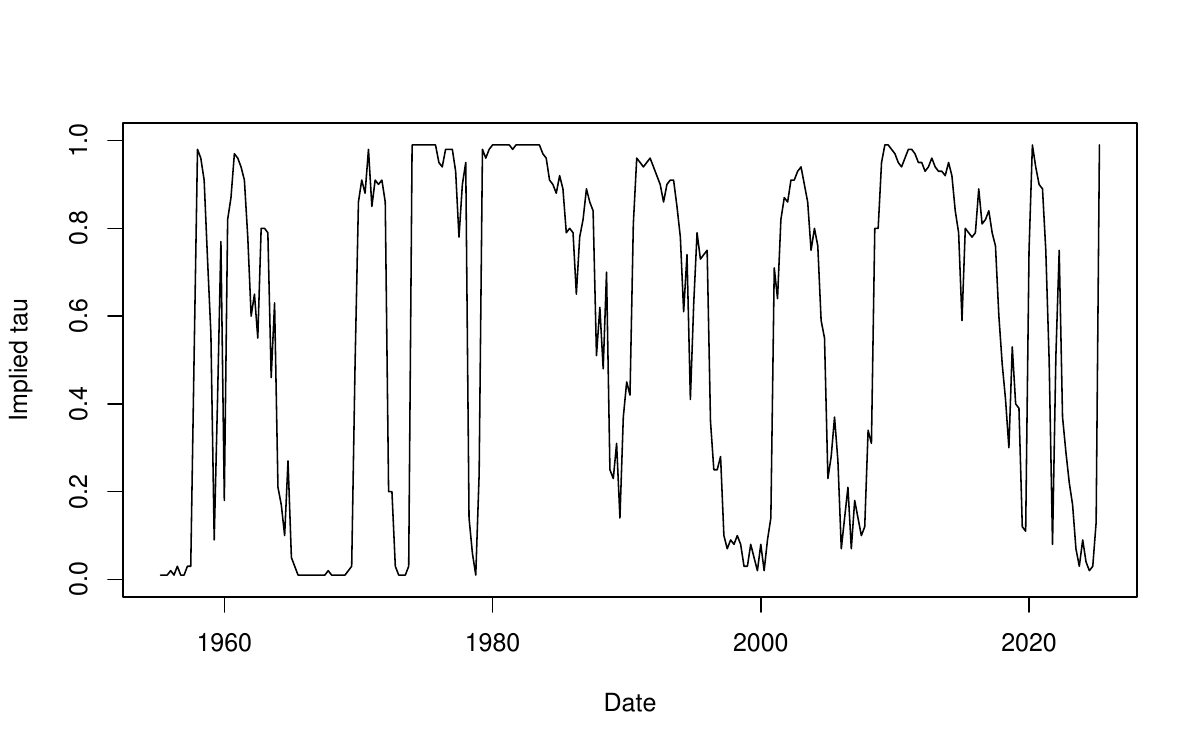}
    \end{subfigure}
 
\end{figure}

\newpage
Figure \ref{fig:baseline:tr} plots the observed Effective Federal Funds Rate (EFFR) together with the optimal interest rate response for a QU maximizer CB (i.e. the Fed), for different representative quantile indexes $\tau\in \{0.1,0.25,0.5,0.75,0.9\}$. The graph represents the wide variety of optimal conditional reactions that may arise for any given QU preference. 

Figure \ref{fig:baseline:tau} plots the implied $\tau$ that represent the closest match to the observed interest rate using a discrete grid search $\tau\in \{0.01,0.02,...,0.98,0.99\}$. Overall, the results indicate that most of the time the Fed has an implied behavior that is consistent with high values of $\tau$. However, in some periods, the implied $\tau$ is drastically reduced.

\begin{figure}[ht]
    \centering
    \caption{Results with dummies: Taylor interest-rate rule and Fed risk aversion.}\begin{subfigure}{0.49\textwidth}
        \centering
       \caption{Observed rate and Taylor rule}
        \label{fig:dummies:tr}        \includegraphics[width=\linewidth]{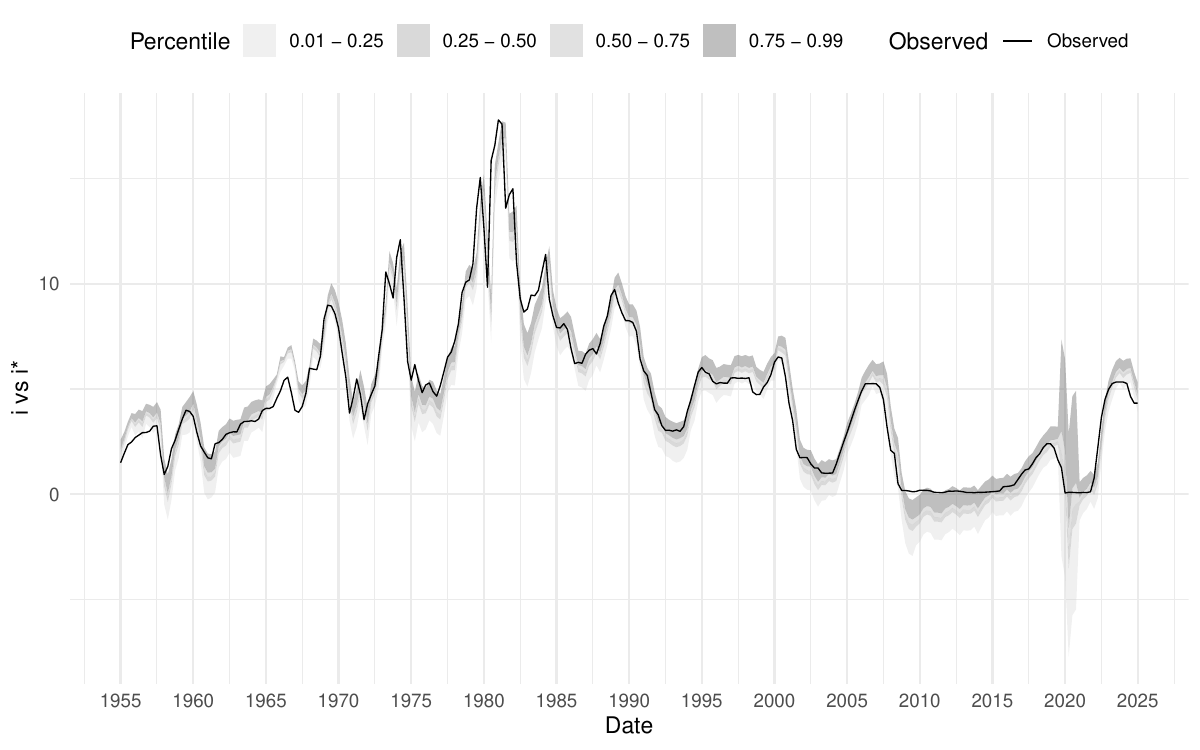}
    \end{subfigure}\hfill
    \begin{subfigure}{0.49\textwidth}
        \centering
        \caption{Implied $\tau$}
        \label{fig:dummies:tau}        \includegraphics[width=\linewidth]{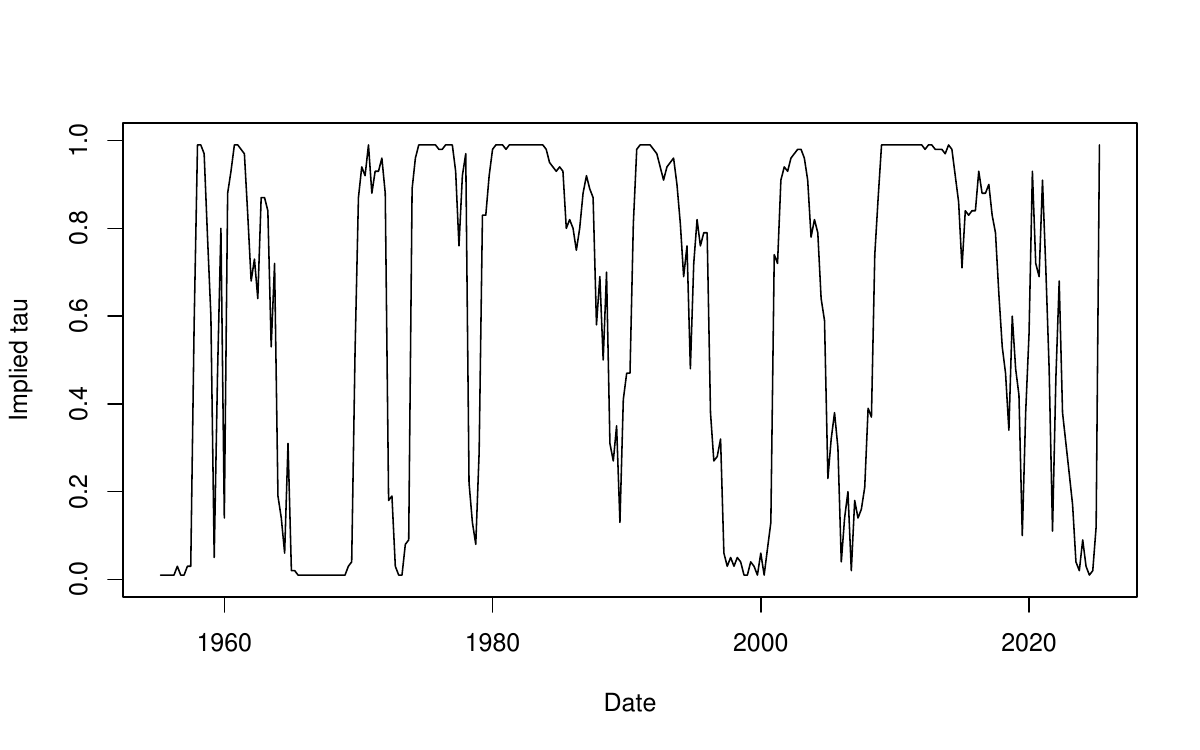}
    \end{subfigure}
    \label{fig:dummies}
\end{figure}

Figures \ref{fig:dummies:tr} and \ref{fig:dummies:tau} presents the same exercise for the model with GFC and COVID dummies. Note that the results are very similar to the baseline model, thus suggesting that the periods highlighted by the dummies are not driving the main results.

In economic terms, a higher $\tau$ means a relatively lower risk aversion by the Fed's authorities. In turn, a lower risk aversion relates to an implied QU preference of the Fed that gives more weight to good macroeconomic outcomes. These findings are consistent with the empirical evidence that shows a significant reduction in inflation and output volatility during the Great Moderation period (see \citeasnoun{stockandwatson2012}, \citeasnoun{bernanke2004}). By contrast, relatively lower values of $\tau$ means higher risk aversion from monetary policy authorities and more weight on potential QU losses. 

Here are some significant episodes where the Fed significantly increased the EFFR. These are in general matched with lower $\tau$ values and changes in the US policymakers behavior regarding their higher risk aversion to undesirable macroeconomic outcomes. More precisely, with the adoption of more hawkish stances: 

1. The Great Inflation Battles (late 1960s-early 1980s). This period was defined by the Fed's struggle against persistently high inflation, which culminated in the most aggressive rate hikes in its history. The Fed began raising rates to combat inflation from the Vietnam War and the 1973 OPEC oil embargo. The EFFR rose from around 3.5 per cent in 1972 to a peak of near 13 per cent in July 1974. From 1977-1980, the inflation rate surged into double digits. Under Fed Chair G. William Miller and then Paul Volcker, the Fed became increasingly aggressive. The rate jumped from approximately 7 per cent in early 1977 to a staggering peak of 20 per cent in April 1980. After a brief easing, inflation remained high. Paul Volcker famously engineered a severe recession to break the back of inflation for good. The Fed drove the rate from near 10 per cent in mid-1980 to a second peak of 19 per cent in June 1981. This is the most famous and aggressive tightening cycle in Fed history.

2. Maintaining Credibility (1983-1984). With inflation now falling, the Fed needed to prove its resolve to keep it down as the economy recovered. The Fed funds rate was raised from near 8.5 per cent to approximately 11.5 per cent to prevent a resurgence of inflation.

3. The Soft-Landing Attempt (1987-1989). With Alan Greenspan now as Chair, the Fed tightened policy as inflation pressures began to build again. The rate rose from near 6.5 per cent to approximately 9.75 per cent.

4. The Preemptive Strike (1994-1995). This is a classic example of a preemptive strike against inflation. The economy was recovering strongly, and the Fed, fearing future inflation, raised rates before inflation actually materialized. A series of rapid hikes took the EFFR from near 3 per cent to approximately 6 per cent. This successful maneuver is often called a soft landing.

5. The Tech Bubble Era (1999-2000). The booming economy and fears of asset bubbles led the Fed to tighten monetary policy. The funds rate was raised from near 4.75 per cent to proximately 6.5 per cent.

6. The Measured Pace (2004-2006). After cutting rates to historic lows of 1 per cent following the dot-com bust, the Fed began a long, predictable cycle of 0.25 per cent hikes to normalize rates. The famous measured pace of 17 consecutive hikes took the funds rate from 1 per cent to 5.25 per cent.

7. The Post-Financial Crisis Liftoff (2015-2018). After seven years near zero following the 2008 crisis, the Fed began a very slow and cautious tightening cycle. Through a series of small, well-telegraphed hikes, the rate moved from near 0.25 per cent to a peak of approximately 2.5 per cent in 2018.

8. The Post-Pandemic Inflation Fight (2022-2023). In response to the highest inflation in 40 years, driven by pandemic stimulus, supply chain issues, and the war in Ukraine, the Fed embarked on its most aggressive tightening cycle since the 1980s. In just over a year, the Fed raised the EFFR from near 0.25 per cent to a target range of 5.25-5.50 per cent, where it remains as of mid-2024.


These findings are revealing in terms of showing that optimal interest rate response by the Fed relates to a specific value of $\tau$ and, accordingly, to a certain degree aversion to undesirable macroeconomic scenarios in terms of inflation and economic activity deviations from their target values. 

Additionally, the implied $\tau$ series also report a consistent behavior with the US monetary policy history observed in the estimation period (see Figure \ref{fig:baseline:tau}). We observe how higher values of $\tau$ and low degrees of risk aversion by the Fed are the norm, with exceptional lower values of $\tau$ and higher degrees of risk aversion related to macroeconomic critical and known events (for a review of the history of monetary policy rules in the US, see \citeasnoun{taylor99}).  

These empirical findings allow us to remark some novel evidence. The dovish or hawkish Fed's behavior varies through time according to the economic, institutional and political context prevailing in the US (see \citeasnoun{eijffingerandmasciandaro2018}). Although we observe that during the great part of estimation period (1959-2025), the Fed's monetary policy authorities display a conduct that approximates more to dovish patterns, we also notice that in critical circumstances, the Fed shows a hawkish stance. 


Our theoretical and empirical contributions differ from the existent ones in the following terms. We introduce a more flexible theoretical framework that maps the dovish/hawkish stances of monetary policy regarding not only the variations of interest rates. Indeed, we consider a more complex analytical framework which allows to define an undesirable macroeconomic scenario in terms of inflation and output deviations from optimal targets jointly. 


\subsection{Robustness and sensitivity analysis}\label{sec:robustness}

\begin{figure}[!htbp]
  \centering
  \captionsetup{justification=centering}
  \caption{Robustness and sensitivity analysis}
  \label{fig:robustness}

  \captionsetup[sub]{justification=centering, position=top, singlelinecheck=false}

  \begin{subfigure}[t]{0.48\textwidth}
    \caption{Taylor rule - $\lambda=0.5$ - Baseline}
    \label{fig:tr_l05_fs}
    \includegraphics[width=\textwidth]{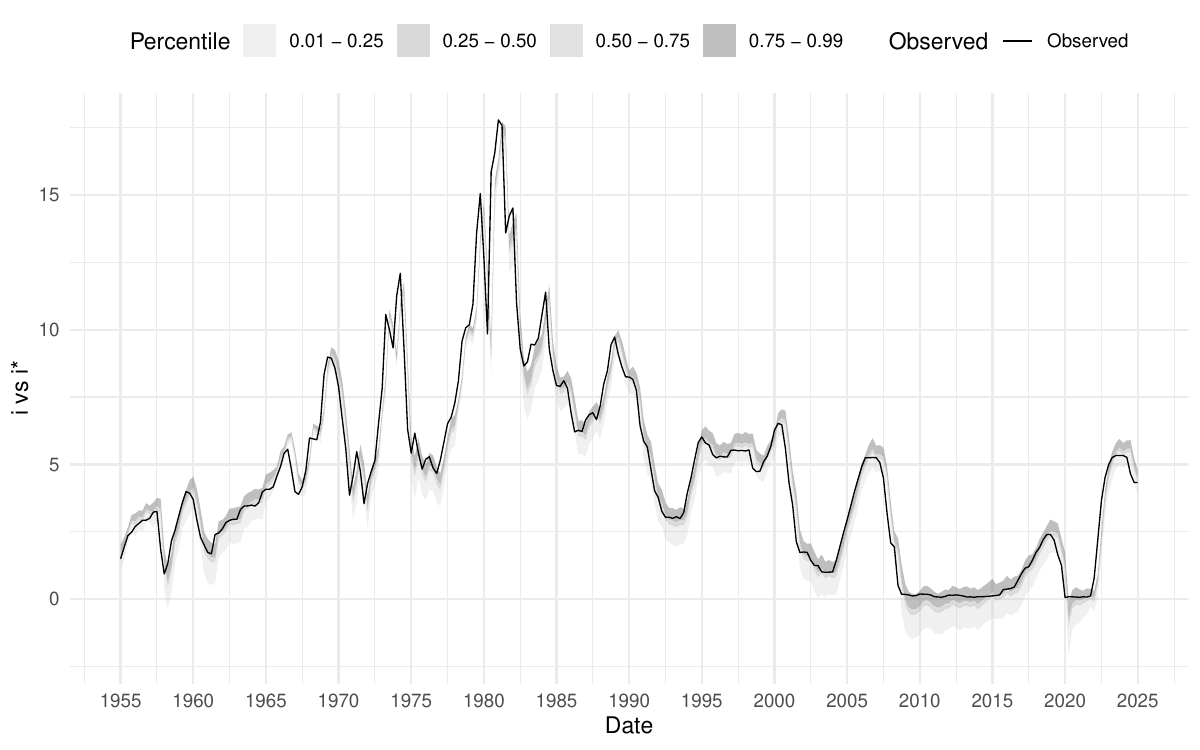}
  \end{subfigure}\hfill
  \begin{subfigure}[t]{0.48\textwidth}
    \caption{Implied $\tau$ - $\lambda=0.5$  - Baseline}
    \label{fig:tau_l05_fs}
    \includegraphics[width=\textwidth]{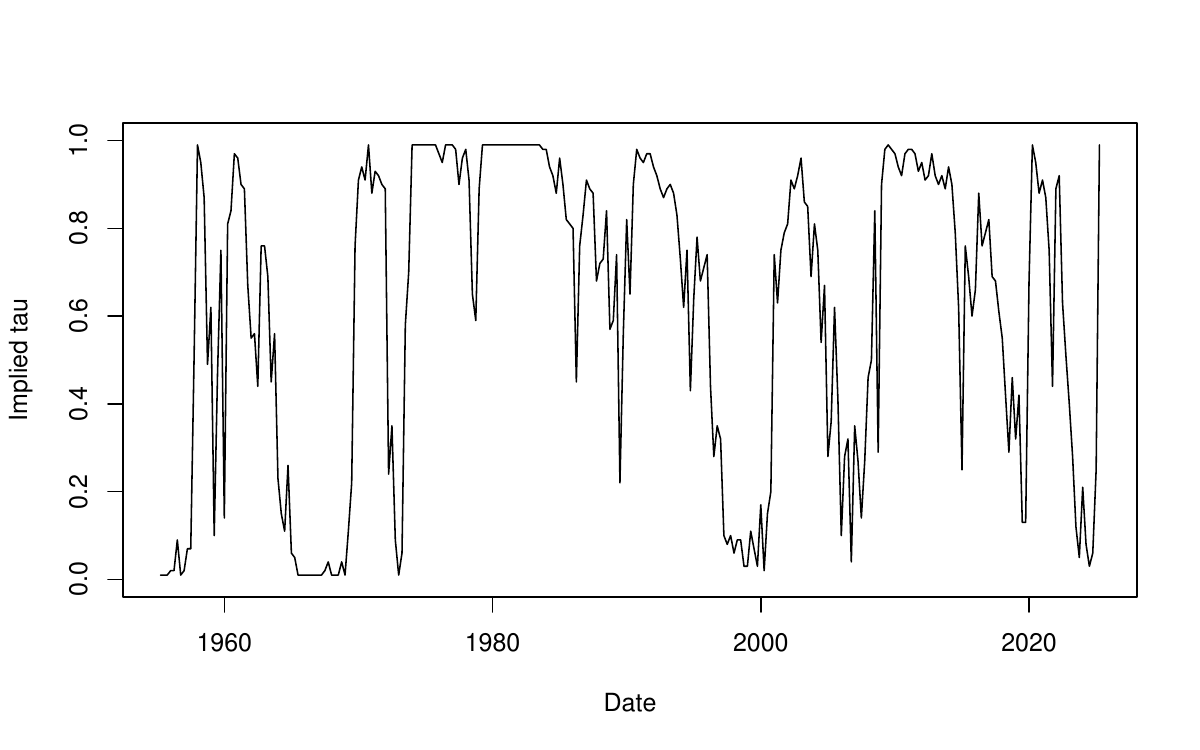}
  \end{subfigure}

  \medskip

  \begin{subfigure}[t]{0.48\textwidth}
    \caption{Taylor rule - $\lambda=2$ - Baseline}
    \label{fig:tr_l2_fs}
    \includegraphics[width=\textwidth]{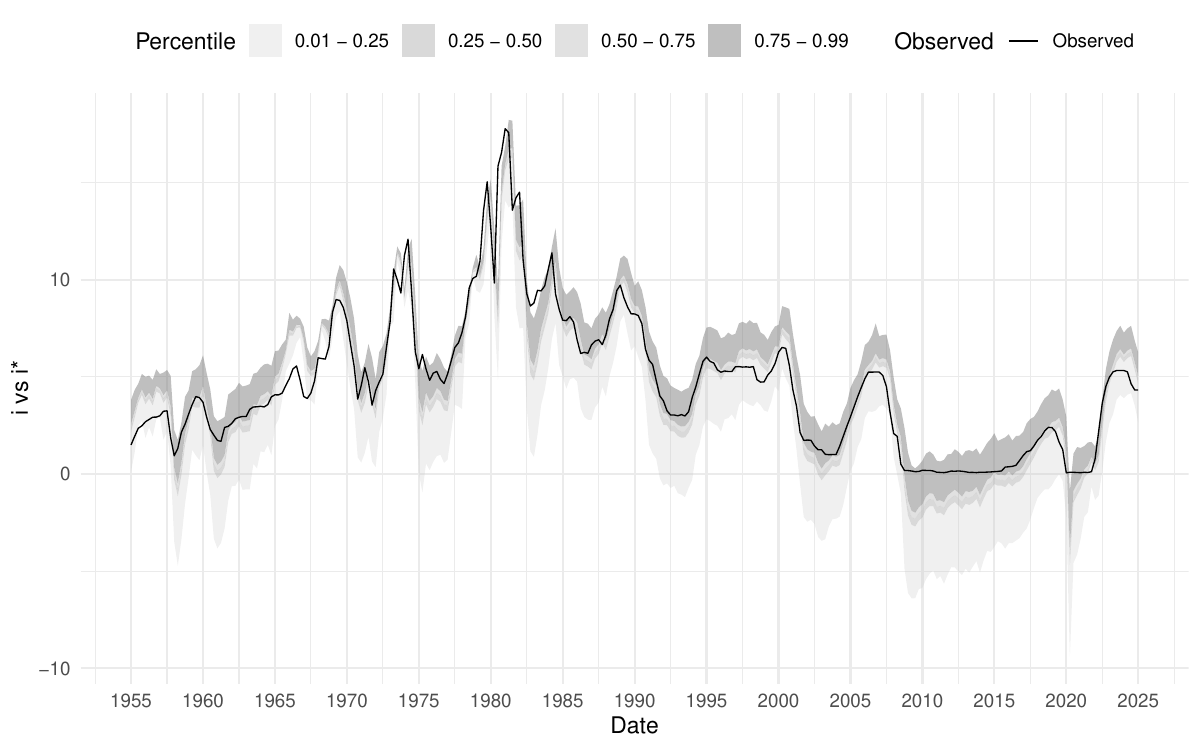}
  \end{subfigure}\hfill
  \begin{subfigure}[t]{0.48\textwidth}
    \caption{Implied $\tau$ - $\lambda=2$- Baseline}
    \label{fig:tau_l2_fs}
    \includegraphics[width=\textwidth]{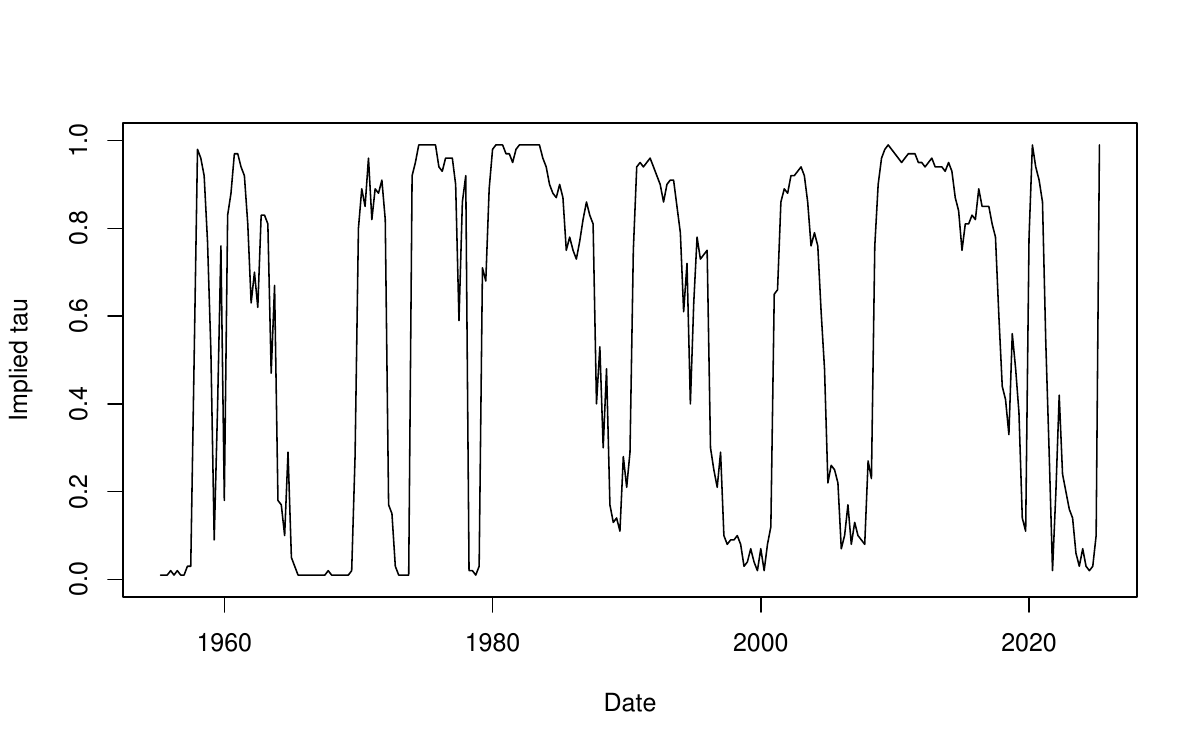}
  \end{subfigure}

  \medskip

  \begin{subfigure}[t]{0.48\textwidth}
    \caption{Taylor rule - $\lambda=1$ - Post-1979}
    \label{fig:tr_l1_p1979_cd}
    \includegraphics[width=\textwidth]{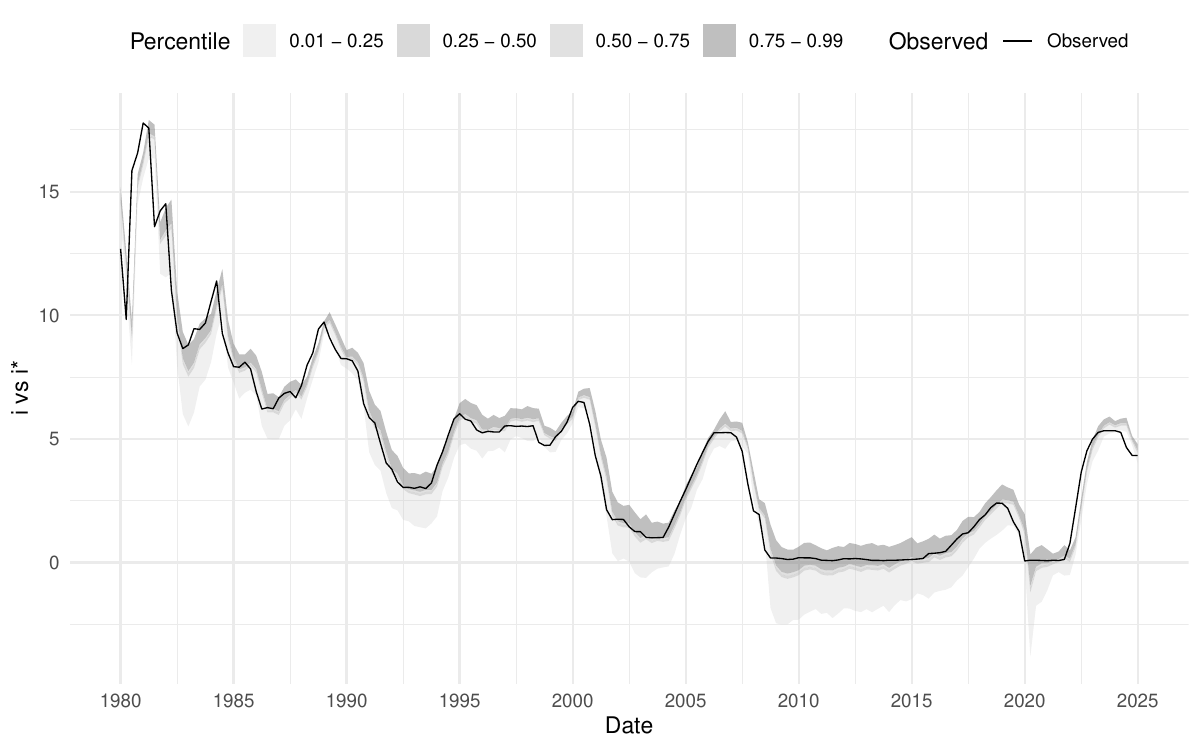}
  \end{subfigure}\hfill
  \begin{subfigure}[t]{0.48\textwidth}
    \caption{Implied $\tau$ - $\lambda=1$ - Post-1979 }
    \label{fig:tau_l1_p1979_sd}
    \includegraphics[width=\textwidth]{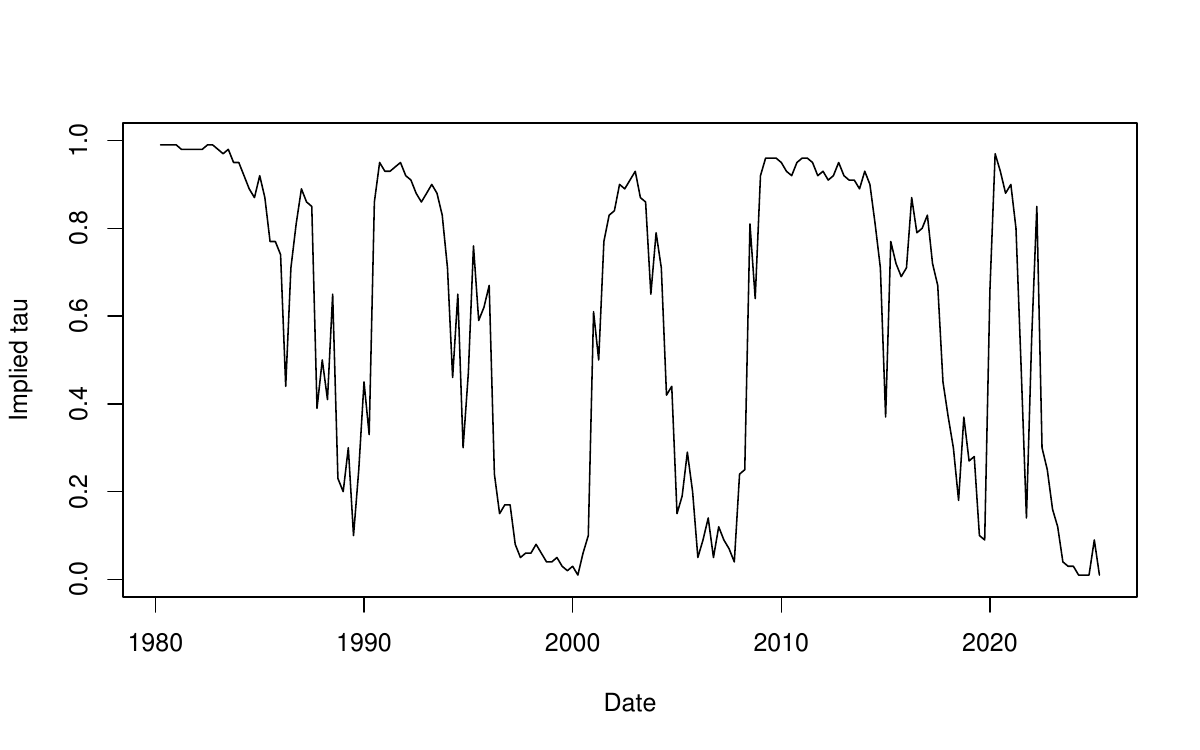}
  \end{subfigure}

  \medskip

  \begin{subfigure}[t]{0.48\textwidth}
    \caption{Taylor rule - $\lambda=1$ - Post-1979 with dummies}
    \label{fig:tr_l1_p1979_sd}
    \includegraphics[width=\textwidth]{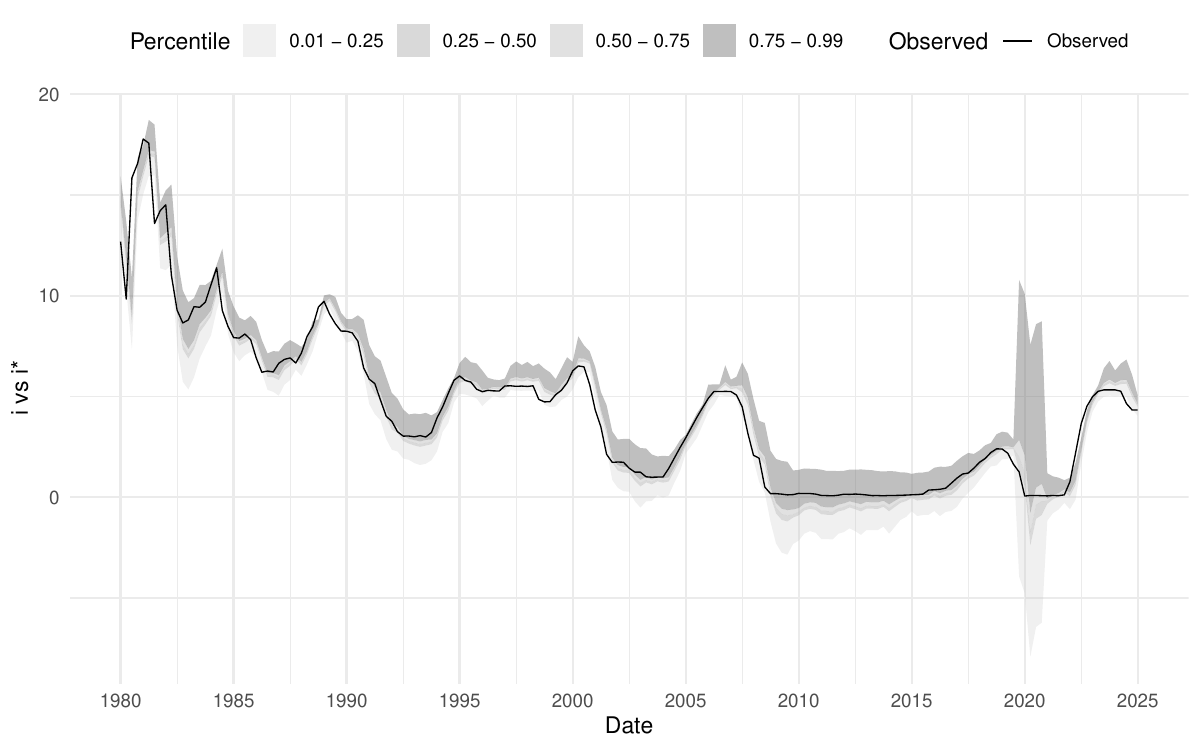}
  \end{subfigure}\hfill
  \begin{subfigure}[t]{0.48\textwidth}
    \caption{Implied $\tau$ - $\lambda=1$ - Post-1979 with dummies}
    \label{fig:tau_l1_p1979_cd}
    \includegraphics[width=\textwidth]{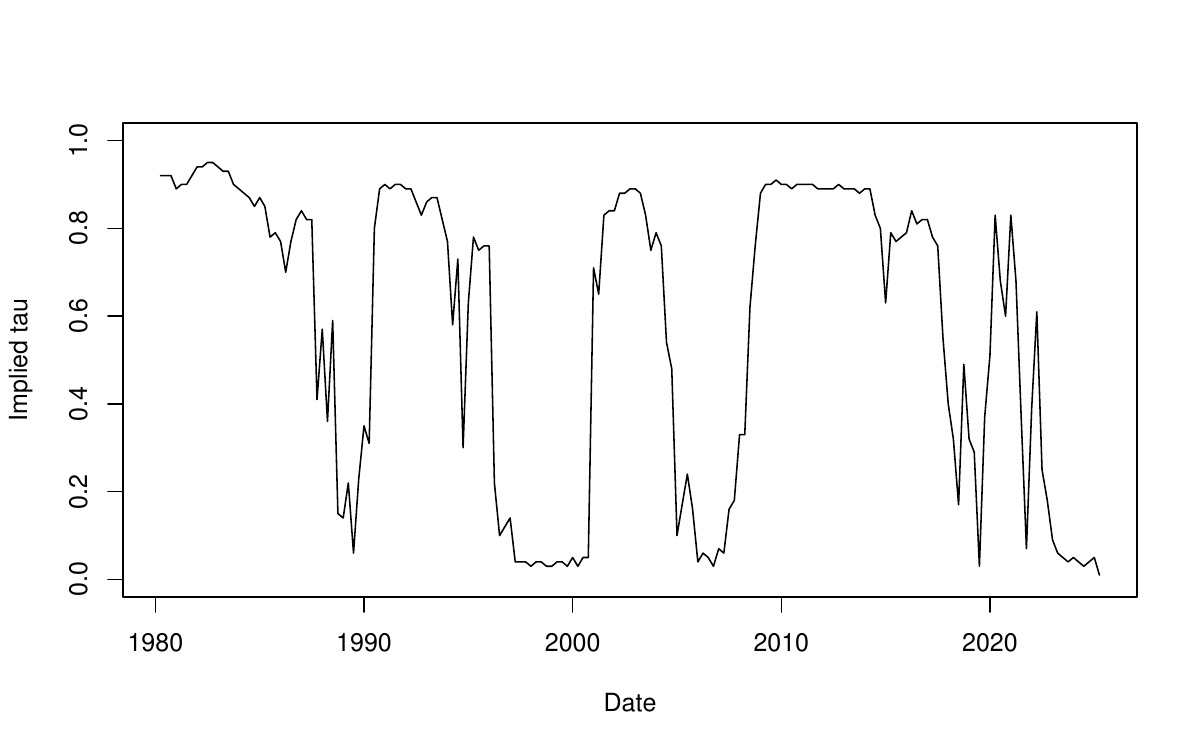}
  \end{subfigure}
\end{figure}

To assess the robustness of our empirical findings, we conducted a battery of sensitivity checks focusing on two main aspects: (i) the relative weight assigned to output stabilization in the CB loss function, parameterized by $\lambda$, and (ii) potential structural breaks in monetary policy behavior associated with the Volcker regime shift.  The results appear in the Tables \ref{tab:var_robustness} and \ref{tab:heteroskedasticity_robustness} and Figure \ref{fig:robustness}.

\begin{table}[!htbp] \centering 
  \caption{VAR(1) Results - Post 1979} 
  \label{tab:var_robustness}
\small
\begin{tabular}{@{\extracolsep{5pt}}lcccc} 
\\[-1.8ex]\hline 
\hline \\[-1.8ex] 
 & \multicolumn{2}{c}{Baseline} & \multicolumn{2}{c}{Baseline + dummies} \\ 
\cline{2-3} \cline{4-5}
 & Inflation ($\pi$) & Output gap ($y$) & Inflation ($\pi$) & Output gap ($y$) \\ 
\\[-1.8ex]\hline \\[-1.8ex] 
$i_{t-1}$ & 0.023$^{***}$ & $-$0.017 & 0.023$^{***}$ & $-$0.026 \\ 
 & (0.008) & (0.023) & (0.009) & (0.023) \\ 
$\pi_{t-1}$ & 0.615$^{***}$ & $-$0.093 & 0.614$^{***}$ & $-$0.128 \\ 
 & (0.061) & (0.165) & (0.061) & (0.162) \\ 
$y_{t-1}$ & $-$0.016 & 0.879$^{***}$ & $-$0.015 & 0.863$^{***}$ \\ 
 & (0.014) & (0.037) & (0.014) & (0.037) \\ 
Constant & 0.144$^{***}$ & 0.054 & 0.145$^{***}$ & 0.177 \\ 
 & (0.048) & (0.130) & (0.050) & (0.133) \\ 
COVID &  &  & 0.074 & $-$0.872$^{*}$ \\ 
 &  &  & (0.171) & (0.452) \\ 
GFC &  &  & $-$0.051 & $-$0.914$^{***}$ \\ 
 &  &  & (0.128) & (0.338) \\ 
\hline \\[-1.8ex] 
Observations & 182 & 182 & 182 & 182 \\ 
R$^{2}$ & 0.545 & 0.760 & 0.546 & 0.773 \\ 
Adjusted R$^{2}$ & 0.537 & 0.756 & 0.533 & 0.766 \\ 
\hline 
\hline \\[-1.8ex] 
\textit{Note:} & \multicolumn{4}{r}{$^{*}$p$<$0.1; $^{**}$p$<$0.05; $^{***}$p$<$0.01} \\ 
\end{tabular} 
\end{table}

\begin{table}[!htbp] \centering 
  \caption{Skedastic Models - Post 1979} 
  \label{tab:heteroskedasticity_robustness} 
\small
\begin{tabular}{@{\extracolsep{5pt}}lcccc} 
\\[-1.8ex]\hline 
\hline \\[-1.8ex] 
 & \multicolumn{2}{c}{Baseline} & \multicolumn{2}{c}{Baseline + dummies} \\ 
\cline{2-3} \cline{4-5}
 & $\hat{u}_\pi^2$ & $\hat{u}_y^2$ & $\hat{u}_\pi^2$ & $\hat{u}_y^2$ \\ 
\\[-1.8ex]\hline \\[-1.8ex] 
$\pi_{t-1}$ & 0.059 & $-$0.514 & 0.104$^{*}$ & 0.131 \\ 
 & (0.065) & (0.781) & (0.059) & (0.560) \\ 
$y_{t-1}$ & $-$0.004 & $-$0.358$^{*}$ & 0.006 & $-$0.216 \\ 
 & (0.017) & (0.209) & (0.016) & (0.150) \\ 
COVID &  &  & 0.294 & 18.803$^{***}$ \\ 
 &  &  & (0.195) & (1.836) \\ 
GFC &  &  & 0.708$^{***}$ & 0.483 \\ 
 &  &  & (0.146) & (1.378) \\ 
Constant & 0.082 & 1.049 & 0.015 & 0.119 \\ 
 & (0.058) & (0.707) & (0.055) & (0.518) \\ 
\hline \\[-1.8ex] 
Observations & 181 & 181 & 181 & 181 \\ 
R$^{2}$ & 0.005 & 0.020 & 0.128 & 0.396 \\ 
Adjusted R$^{2}$ & $-$0.006 & 0.009 & 0.108 & 0.382 \\ 
\hline 
\hline \\[-1.8ex] 
\textit{Note:} & \multicolumn{4}{r}{$^{*}$p$<$0.1; $^{**}$p$<$0.05; $^{***}$p$<$0.01} \\ 
\end{tabular} 
\end{table}

First, we explored the sensitivity of the estimated Taylor-type quantile rule to alternative values of $\lambda \in \{0.5, 1, 2\}$. This exercise allows us to gauge whether the implied policy stance and the inferred degree of risk aversion are contingent upon the assumed trade-off between inflation and output stabilization. The results remained qualitatively stable across specifications. Lower values of $\lambda$ ($0.5$) led to slightly more aggressive responses to inflation deviations, reflecting a relatively more hawkish stance, while higher values ($2$) induced smoother interest rate paths, consistent with greater tolerance toward output fluctuations. Nevertheless, the implied quantile-based preferences ($\tau_t$) maintained the same cyclical pattern, confirming that the estimated policy reaction functions are not overly sensitive to moderate changes in the structural weighting scheme.

Second, to account for potential regime shifts in the conduct of US monetary policy, we re-estimated the baseline VAR and conditional heteroskedasticity models using a truncated sample starting in the fourth quarter of 1979. This subsample captures the onset of the Volcker disinflation episode, a well-documented structural change in the Fed's reaction to inflationary pressures. The results reported show that the estimated coefficients remain broadly consistent with those of the full sample, with minor quantitative adjustments reflecting a stronger disinflationary response of the Fed in the post-Volcker era. The inferred quantile preferences confirm a temporary decline in $\tau_t$ during the early 1980s, signaling a shift towards higher risk aversion and more hawkish policy attitudes, followed by a gradual return to higher $\tau_t$ values consistent with a more dovish stance in subsequent decades.

Overall, both exercises confirm the internal consistency and empirical robustness of our quantile-based Taylor rule. The results suggest that the main findings, the predominance of dovish-type behavior with episodic hawkish responses are not artifacts of parameter calibration or sample selection, but rather reflect persistent structural features of the US monetary policy.

\section{Conclusions and discussion}\label{sec:conclusions}

The study of Taylor rules through quantile methods highlights that a one-size-fits-all linear rule may be inadequate, with more nuanced, quantile-aware approaches needed to understand and formulate policy, especially in diverse economic environments. A CB may not react linearly to economic variables; its responses can vary significantly at different quantiles of the distribution. 

The QU framework allows to study how independent variables (like policy tools) impact the dependent variable (like interest rates) differently across the entire distribution of outcomes, revealing heterogeneity in policy reactions. QU models explore how agents make decisions under uncertainty by focusing on outcomes at different parts of the probability distribution, while Taylor rules describe how the CBs set interest rates based on inflation and economic output. 

In this paper, we study optimal monetary policy when a CB maximizes a QU objective rather than expected utility operator. In our framework, the CB's risk attitude is indexed by the quantile level $tau$, providing a transparent mapping between hawkish/dovish stances and attention to adverse macroeconomic realizations. We formulate the infinite-horizon problem using a Bellman equation with the quantile operator. Implementing an Euler-equation approach, we derive Taylor-rule-type reaction functions. The Taylor rule is recovered as a special case when quantiles replace the expectation operator. An empirical implementation is outlined based on reduced-form VAR(1) laws of motion with conditional heteroskedasticity, enabling estimation of the new rule and its dependence on risk attitudes. It is important to note that our analytical methodology makes the assumption that the policymaker's statements are entirely credible. Determining optimality criteria for establishing a Taylor rule inside the quintile preference framework in the setting of monetary policy pronouncements lacking credibility is a tenable extension of this study.

If the CB is relatively more risk averse, it would react more aggressively to downside risks than to equivalent upside risks. If the CB is relatively less risk averse, policies are guided by the possibility of good economic results. This could justify the risk management approach often discussed at CBs. 

Quantile preferences naturally incorporate concerns about tail risks that expected utility might underweight. Moreover, risk attitudes might not be constant across time. Our model allows us to identify risk aversion behavior through indirect inference, by mapping the observed policy variables with the corresponding value in the optimal CB behavior.

Our empirical results for the US show that the Fed has mostly a dovish attitude over long periods of time (higher $\tau$), but with hawkish attitudes in specific periods (lower  $\tau$). These periods coincide with regime changing events, like the oil crisis in the middle of the 1970s, the Volcker new approach towards fighting inflation at the end of the 1970s, the global financial crash of 2008, and the post COVID pandemia. As such, the implied risk aversion estimates reveal important changes in the Fed preferences.

The policy implications are that focusing solely on the mean in economic modeling, as many traditional Taylor rules applications do, can be misleading, as non-linear relationships and risk preferences (represented by QU) can significantly alter optimal monetary policy and under specific conditions. 

\bibliographystyle{econometrica}
\bibliography{quantile}


\end{document}